\pdfoutput=1

\documentclass[preprint,12pt]{elsarticle}

\usepackage[utf8]{inputenc}
\usepackage{xcolor}

\raggedright
\usepackage{amsmath, amsfonts, amssymb, amsthm, mathtools, commath}
\usepackage{array, booktabs, makecell, mhchem}
\usepackage{mathpazo, mathrsfs}
\usepackage{bm}
\usepackage{booktabs}
\usepackage{textcomp}
\usepackage{siunitx}
\usepackage[aboveskip=1pt,labelfont=bf,labelsep=period,justification=raggedright,singlelinecheck=false, tableposition=top]{caption}
\usepackage[labelformat=empty,position=top]{subcaption}
\usepackage[export]{adjustbox}
\usepackage{siunitx,booktabs,threeparttable}
\usepackage{floatrow,float}
\newfloatcommand{capbtabbox}{table}[][\FBwidth]

\usepackage{nameref,hyperref}

\usepackage{siunitx, booktabs,threeparttable}
\usepackage{tabularx}
\usepackage{multirow}
\usepackage{chngcntr}
\usepackage{stackengine}

\usepackage[english]{babel}
\usepackage{textcomp,marvosym}

\DeclareMathOperator{\ELBO}{ELBO}
\DeclareMathOperator{\EX}{\mathbb{E}}

\usepackage{xr}
\usepackage{cleveref}

\journal{arXiv}

\begin{document}

\begin{frontmatter}



\title{Technical outlier detection via convolutional variational autoencoder for the ADMANI breast mammogram dataset}


\author[aff1]{Hui Li}
\author[aff2]{Carlos A. Peña-Solorzano}
\author[aff1]{Susan Wei}
\author[aff1,aff2,aff3]{Davis J. McCarthy}

\affiliation[aff1]{organization={School of Mathematics and Statistics, University of Melbourne},%Department and Organization
            addressline={813 Swanston Street, Parkville}, 
            city={Melbourne},
            postcode={3010}, 
            state={VIC},
            country={Australia}}

\affiliation[aff2]{organization={Bioinformatics and Cellular Genomics, St Vincent's Institute of Medical Research},%Department and Organization
            addressline={9 Princes St, Fitzroy}, 
            city={Melbourne},
            postcode={3065}, 
            state={VIC},
            country={Australia}}

\affiliation[aff3]{organization={Melbourne Integrative Genomics, University of Melbourne},%Department and Organization
            addressline={Building 184/30 Royal Parade, Parkville}, 
            city={Melbourne},
            postcode={3052}, 
            state={VIC},
            country={Australia}}

\begin{highlights}
\item Convolutional variational autoencoder (CVAE), supplemented with traditional image processing techniques (i.e. erosion, pectoral muscle analysis), can detect a variety of technical outliers present in the ADMANI dataset.
\end{highlights}

\begin{abstract}
The ADMANI datasets (annotated digital mammograms and associated non-image datasets) from the ``Transforming Breast Cancer Screening with AI" programme (BRAIx) run by BreastScreen Victoria in Australia are multi-centre, large scale, clinically curated, real-world databases. The datasets are expected to aid in the development of clinically relevant Artificial Intelligence (AI) algorithms for breast cancer detection, early diagnosis, and other applications. To ensure high data quality, technical outliers must be removed before any downstream algorithm development. As a first step, we randomly select 30,000 individual mammograms and use Convolutional Variational Autoencoder (CVAE), a deep generative neural network, to detect outliers. CVAE is expected to detect all sorts of outliers, although its detection performance differs among different types of outliers. Traditional image processing techniques such as erosion and pectoral muscle analysis can compensate for CVAE's poor performance in certain outlier types. We identify seven types of technical outliers: implant, pacemaker, cardiac loop recorder, improper radiography, atypical lesion/calcification, incorrect exposure parameter and improper placement. The outlier recall rate for the test set is $61\%$ if CVAE, erosion and pectoral muscle analysis each select the top $1\%$ images ranked in ascending or descending order according to image outlier score under each detection method, and $83\%$ if each selects the top $5\%$ images. This study offers an overview of technical outliers in the ADMANI dataset and suggests future directions to improve outlier detection effectiveness.
\end{abstract}



\end{frontmatter}


\section{Introduction \label{sec:introduction}}
During breast cancer screening and diagnosis, mammogram images should meet stringent technical image quality criteria \cite{perry2008european, van2015right}. Technical artefacts reduce image quality, imitate or mask clinical abnormalities, and cause interpretation errors \cite{geiser2011challenges}. For example, blurring artefacts caused by patient movement during mammogram scanning obscures microcalcifications \cite{ayyala2008digital}; some types of antiperspirant and skin creams used by patients appear as high-density particles on mammograms and have a similar appearance to microcalcifications or unusual lesions \cite{geiser2011challenges}; implanted medical devices can obscure part of mammogram, reduce contrast in craniocaudal (CC) and mediolateral–oblique (MLO) views, and reduce projected breast tissue and pectoral muscle projection \cite{paap2016mammography}; whereas misread or dead pixels due to detector problems appear as clusters of microcalcifications \cite{geiser2011challenges, ayyala2008digital}. As a result, it is critical to detect and remove mammograms with technical artefacts prior to any downstream diagnosis.

Mammograms containing technical artefacts are considered outliers or anomalies when compared to mammograms without such artefacts. The primary purpose of outlier detection is to identify data points that differ significantly from the rest and are suspected to be the product of a separate cause \cite{hawkins1980identification}. Outlier detection is challenging due to outliers' irregular behaviour and the lack of uniform criteria for what constitutes an outlier \cite{chalapathy2019deep, fernando2020neural, fernando2021deep}. For example, in general mammograms of malignant breast are considered as abnormal compared with mammograms of healthy breast. However, both cancerous and healthy breast mammograms are considered as normal when abnormal are defined as those with technical artefacts in this research.

Breast mammogram datasets are large-scale and each image is high-dimensional, which adds to the task of detecting outliers. Because of the size of the data set, outlier detection algorithms are required to be both fast and scalable \cite{koufakou2009scalable, thudumu2020comprehensive}. The ``curse of dimensionality" issue for high-dimensional data causes distance concentration and the creation of hubs and antihubs \cite{radovanovic2010hubs, radovanovic2014reverse}, which makes outlier discovery more difficult. To be more specific, it is widely understood that as dimension grows, distances between pairwise points become progressively indistinguishable (distance concentration) and every point is detected as a nearly equally good outlier, even if there are inliers and outliers. Some research challenges this viewpoint, claiming that particular sites (such as antihubs) are more likely to be discovered as outliers even when no actual outliers are expected.

%``curse of dimensionality" includes various phenomena: increased data sparsity (i.e. the high-dimensional data points are more scattered and isolated due to the unnecessary attributes or high noise level of multiple irrelevant covariates) \cite{bellman1966dynamic, thudumu2020comprehensive}, distance concentration (i.e. the distances of all the data points to some arbitrary reference point are indiscernible since the ratio of the constant standard deviation to the increasing expected mean of the distribution of the distances converges to 0 when dimension increases to infinity) \cite{radovanovic2014reverse, radovanovic2010hubs}, emergence of hubs and antihubs (i.e. hubs refer to the points which are relatively closer to the reference point and occurs more frequently among the k-Nearest Neighbors of all other data points in the data set when dimension increases, while antihubs are just the opposite) \cite{radovanovic2014reverse, radovanovic2010hubs}. 

%Therefore, for outlier detection, as dimension increases, it is commonly accepted that distances between pairwise points are increasingly indiscernible (i.e. distance concentration) and every point is detected as an almost equally good outlier. This view is challenged by some research \cite{radovanovic2014reverse} since some points (i.e. antihubs) are more prominent to be detected as outliers even when no true outliers are expected. Both the distance concentration and antihubs contribute to the difficulty in outlier detection for high-dimensional dataset.

Deep learning is particularly suited to large scale and high dimensional tasks. During training, the large-scale dataset is split into mini-batches and fed into the deep neural networks (DNN) for optimization. To tackle the ``curse of dimensionality" issue, the high dimensional data is reduced into low-dimensional representations in the latent space of the DNN. Compared to traditional dimensionality reduction algorithms such as principal component analysis, DNN is able to learn more complex data patterns due to the hierarchical and non-linear mapping from input to latent representations. The ability to learn complex patterns is especially important for medical datasets where the data points are inherently heterogeneous or noisy. In the ADMANI databases \cite{frazer2022admani}, breast mammograms with various types of anomalous technical artefacts are considered outliers, whereas inliers comprise both healthy and malignant breasts. The continuous development from healthy to cancer state, as well as the many cancer sub-types result in a highly heterogeneous composition for inliers, making modelling the complex normal data distribution more difficult \cite{fernando2021deep}. The complex or computationally intractable data distributions learnt by DNN \cite{lu2020universal} can thus better segregate anomalous data from normal ones. 

Because outlier labels are rarely known in advance, unsupervised deep learning is commonly utilized in outlier detection. Autoencoders (AEs), Variational Autoencoders (VAEs), Vector Quantised-Variational Autoencoders (VQ-VAEs), Generative Adversarial Networks (GANs), and one-class neural networks (OC-NNs) are some of the most often utilized unsupervised DNNs. Convolutional variational autoencoders (CVAE) are simply VAEs with convolutional layers required to analyze image data.

AEs encode the inputs into latent representations, which are then reconstructed to the original input. Autoencoders and its variants are widely applied in outlier detection \cite{hawkins2002outlier, tagawa2015structured, chalapathy2017robust, chen2018autoencoder}. VAEs are deep generative models with latent variables of which the posterior distribution is parameterized with neural networks. VAEs were proposed as anomaly detection algorithms by An and Cho in 2015 and were demonstrated to outperform AEs in an experimental setting, where anomaly was defined as one class in the MNIST/KDD dataset \cite{an2015variational}. More examples of VAEs being used in anomaly identification can be found in \cite{lu2018anomaly, zimmerer2018context, matias2021robust, xu2018unsupervised}. The reasoning for AE and VAE to detect outliers is that when AE or VAE is trained on a dataset with no or few outliers, during inference time it will be more difficult to reconstruct or model an anomalous input, resulting in larger reconstruction loss\cite{an2015variational}.

VQ-VAEs combine VAEs with vector quantization to obtain a discrete latent representation; the prior of the latent representation is learnt to be auto-regressive \cite{van2017neural}. The expressive auto-regressive models of the latent representation enables state-of-art density estimation in images, and has been demonstrated to outperform VAEs in both sample-wise and pixel-wise anomaly detection when applied to brain MR and abdominal CT datasets \cite{marimont2021anomaly}. GAN-based anomaly detectors are extensively summarized in the review by Xia et. al. in 2022 \cite{xia2022gan}. Generally speaking, GAN-based anomaly detection methods use normal data points to train a feature extractor which maps the data to its latent representations. During inference time, a latent representation is predicted for the test data point and an anomaly is determined by comparing the reconstructed and the test data point through the generator \cite{li2018anomaly}, discriminator \cite{donahue2016adversarial} or both \cite{schlegl2017unsupervised}. In contrast with AE, VAE, VQ-VAE, and GAN where the latent representations are tailored for reconstruction first and then applied to traditional anomaly detection methods, OC-NN was proposed \cite{chalapathy2018anomaly,ruff2018deep} in which the latent representations are trained by maximising an objective function tuned for anomaly detection directly. 

The anomalies discovered in the mammography scans that comprise the ADMANI dataset are the focus of this study. We chose CVAE as the model after comprehensive considerations of outlier identification performance and neural network architectural complexity. Although VAE can target different types of outliers with varying success rates, the performance was supplemented using simple hand-crafted outlier identification approaches that target specific kinds of outliers.

\section{Method}
\subsection{Dataset}
The ADMANI (annotated digital mammograms and associated non-image) datasets from the  ``Transforming Breast Cancer Screening with AI" programme (BRAIx) run by BreastScreen Victoria in Australia are investigated in this research. As the current largest breast mammogram dataset in the world \cite{frazer2022admani}, ADMANI datasets are 1) large scale with specifically 4,410,212 breast mammogram images from 629,700 clients, 2) longitudinal spanning from 2013 to 2019 with intention to grow continually in subsequent years, 3) enhanced with associated client demographic and clinical non-image data. The large clinically curated real-world ADMANI datasets are established to aid in the development of clinically relevant Artificial Intelligence (AI) algorithms for breast cancer detection, early diagnosis, and other applications. Removing images with technical artefacts is necessary to ensure high data quality for any downstream AI algorithm development. We randomly choose 30,000 mammogram images with their related meta data for the technical outlier detection.

\subsection{CVAE \label{sec:CVAE}}
The image dataset can be represented by an $N \times C \times H \times W$ matrix $\bm X$, where $N$ is the number of images, $C$ is the number of channels per image, and $H$ and $W$ are the image height and width, respectively. Since the mammograms present in the ADMANI dataset are gray-scale, $C = 1$, and will be omitted as matrix dimension on latter sections. For each image $\bm{x}_{i}$, where $i=1,...,N$, the entry $\bm{x}_i^{h \times w}$ refers to the intensity value of the pixel at row $h$ and column $w$, where $h=1,...,H$ and $w=1,...,W$. The generative model for $\bm{x}_i$ is $p(\bm{x}_i|\bm{z}_i)$, where $p$ is a multivariate Gaussian distribution with parameters $\bm{\mu}_i$ and $\bm{\sigma}_i$, and $\bm z_i$ is a low-dimensional latent vector for $\bm x_i$ with a prior distribution $p(\bm z_i) = \mathcal{N}(\bm 0, \bm I)$, where $I$ is the identity matrix. The generative model is shown in Eq. \eqref{eq:generativemodel}.
\begin{equation}
\begin{split}
\bm z_i & \sim \mathcal{N}(\mathbf{0},\mathbf{I})\\
\bm \mu_i & = f_{\bm \theta}(\bm z_i)\\
\bm x_{i} & \sim \mathcal{N}(\bm \mu_{i}, \bm \sigma_{i}),
\end{split}
\label{eq:generativemodel}
\end{equation}
where $f_{\bm \theta}$ is a deconvolutional neural network (decoder) with parameters $\bm \theta$, mean $\bm \mu_i$ ($H \times W$ vector), and variance parameter $\bm \sigma$, which is assumed to be the $H \times W$ identity matrix. The spatial location of each pixel in image $\bm x_i$ is recovered by the deconvolution process.

The distribution of the posteriors $p(\bm z|\bm x)$ is intractable and could be approximated by Variational inference (VI), a faster alternative to Markov chain Monte Carlo (MCMC). Variational inference is a process to find a member distribution $q$ in a predefined approximated distribution family $Q$ to satisfy some criteria, and $q \in Q$. For mathematical convenience, the approximated posterior family $Q$ is assumed to be a multivariate Gaussian family with diagonal covariance matrix. Since the mean and variance parameters of $q$ are learnt through an encoder convolutional neural network with weights $\bm \phi$ and input $\bm x$, the distribution $q$ is further denoted as $q_{\bm \phi}(\bm z| \bm x)$. The criteria of VI is to find a member $q_{\bm \phi}(\bm z|\bm x)$ which minimizes the Kullback-Leibler (KL) divergence (D) between $q_{\bm \phi}(\bm z | \bm x)$ and the intractable true posterior $p(\bm z | \bm x)$.

\begin{equation}
D(q_{\bm \phi}(\bm z |\bm x)||p(\bm z |\bm x))=\EX_{q_{\bm \phi}(\bm z |\bm x)}(\log q_{\bm \phi}(\bm z |\bm x) - \log p(\bm z |\bm x)).
\label{eq:KL}
\end{equation}

After applying the Bayes rule and rearranging, Eq. \eqref{eq:KL} can be rewritten as:

\begin{equation}
\begin{split}
 \log p(\bm x) - D(q_{\bm \phi}(\bm z|\bm x)||p(\bm z|\bm x)) &= \EX_{q_{\bm \phi}(\bm z|\bm x)}\log p_{\bm \theta}(\bm x|\bm z) - D(q_{\bm \phi}(\bm z|\bm x)||p(\bm z)), 
\end{split}
\end{equation}

where $p(\bm z)$ is the prior. Since $D(q_{\bm \phi}(\bm z|\bm x)||p(\bm z|\bm x))$ is non-negative, we have

\begin{equation}
 \log p(\bm x) \geq \EX_{q_{\bm \phi}(\bm z|\bm x)}\log p_{\bm \theta}(\bm x|\bm z) - D(q_{\bm \phi}(\bm z|\bm x)||p(\bm z)).  
 \label{eq:lower_bound}
\end{equation}

The right side of the Eq. \eqref{eq:lower_bound} is called the Evidence Lower Bound (ELBO) for the log evidence, $\log p(\bm x)$. Minimizing the KL divergence in Eq. \eqref{eq:KL} is equivalent to maximizing the ELBO. Given a training set $\{(\bm x_i)\}_{i=1}^{N}$, define

\begin{equation}
\begin{split}
    \ELBO(\bm \phi, \bm \theta) = \frac{1}{N}\sum_{i=1}^{N} \EX_{q_{\bm \phi}(\bm z_i|\bm x_i)} \log p_{\bm \theta}(\bm x_i|\bm z_i)- \frac{1}{N}\sum_{i=1}^{N} D(q_{\bm \phi}(\bm z_i|\bm x_i) || p(\bm z_i)).
\end{split}
\label{eq:ELBO}
\end{equation}

The plug-in estimator of the expectation in the reconstruction term is for $L$ samples of $\bm z$ from the posterior distribution $q_{\bm \phi}(\bm z_i|\bm x_i)$ for each $\bm x_i$. As a standard practice, the hyperparameter $L$ is set to be 1. Taking together that $p_{\bm \theta}(\bm x_i|\bm z_i)$ is a multivariate Gaussian with identity matrix and the analytical form of KL-divergence is available when one probability distribution is Gaussian with diagonal covariance and the other is standard Gaussian, we have
\begin{equation}
\begin{split}
    \ELBO(\bm \phi, \bm \theta) =& - \frac{H \times W}{2}\log (2\pi) + \frac{1}{N}\sum_{i=1}^{N}\left\{\sum_{t=1}^{H \times W}-\frac{1}{2}\big(\bm x_{ijt} - (f_{\bm \theta}(\bm z_{ij})_t \big)^2\right\}\\
    &+ \frac{1}{N}\sum_{i=1}^{N}\left\{- \frac{1}{2}\sum_{j=1}^{K}\big(\sigma_{ij}^2 + \mu_{ij}^2 - 1 - \ln (\sigma_{ij}^2)\big)\right\},
\end{split}
\label{eq:ELBO_outlier}
\end{equation}
where $H$ and $W$ are image height and width, $K$ is the dimension for latent embeddings, and $\mu_{ij}$ and $\sigma_{ij}$ are mean and variance for dimension $j$ of latent vector $\bm z_i$ with posterior $q(\bm z_i|\bm x_i)$. The expression $- \frac{H \times W}{2}\log (2\pi)$ is constant and can be omitted. The second term of ELBO in Eq. \ref{eq:ELBO_outlier} is referred to as negative Reconstruction loss, while the third term is referred to as negative KLD loss. 

\subsection{CVAE at various depths}
\label{sec:CVAE_varying_depths}
We compared two CVAE architectures, VanillaCVAE \cite{kingma2013auto} and ResNetCVAE \cite{julianstastny}. VanillaCVAE is the architecture when VAE was first introduced by Kingma \textit{et.al.} in 2013 and it only has 5 layers for both encoder and decoder. ResNetCVAE is deeper than VanillaCVAE with 18 layers for both encoder and decoder. To avoid degradation of training accuracy for the deep ResNetCVAE, the ordinary layers are reformulated as learning residual functions with reference to the layer inputs by adding identity shortcut connections \cite{he2016deep}.

For both VanillaCVAE and ResNetCVAE, we used grid-search among 24 alternatives to get the best configuration. These 24 configuration options are determined empirically. Because greater ELBO is not a consistent criterion across different setups, the ideal hyper-parameters were chosen by visually inspecting the generated output image. Table \ref{tab:tune_hyperparameters} shows the grid values and the ideal configuration that was chosen. For all the 24 configurations, the ADMANI dataset was split into train, valid and test at a ratio 0.6:0.1:0.3 and the neural network was trained with Adam optimiser, a learning rate 0.0005, batch size 64 for 100 epochs. The generated image is deemed to be with satisfactory quality for the optimal VanillaCVAE and ResNetCVAE selections (see Figure S1 in supplementary file).

\begin{table}[h!]
\resizebox{\textwidth}{!}{%
  \begin{tabular}{@{} lcccc @{}}
    \toprule
     & \textbf{resize height} & \textbf{resize width} & \textbf{number of output channels} & \textbf{latent dimension}\\
    \midrule
    Grid values & 256, 512 & 256 & 8, 16, 32 & 128, 256, 512, 1024 \\
    Optimal VanillaCVAE & 512 & 256 & 8 & 512 \\
    Optimal ResNetCVAE & 256 & 256 & 8 & 256\\
    \bottomrule
  \end{tabular}%
}%
\caption{\textbf{Tune hyperparameters for VanillaCVAE and ResNetCVAE.} Resize means the resize transformation for the input image. The number of output channels is for the output from the first hidden layer in encoder. The number of input channels for the first layer in encoder is 1 since the breast mammogram is grayscale. Latent dimension is the dimension of the latent representation for each image. \label{tab:tune_hyperparameters}}
\end{table}

\subsection{Outlier scores \label{sec:outlier_scores}}
We defined 15 outlier scores, which are grouped into three categories: 1) associated with the generative loss, 2) associated with the latent representation, and 3) a mix of the generative loss and the latent representation. For all of the 15 outlier scores, smaller values indicate outliers and are assigned numerals for ease of reference, as shown in Table~\ref{tab:outlier_score_abbrev}. The following section provides a detailed explanation of the outlier scores.

\begin{table}[h!]
\resizebox{\textwidth}{!}{%
  \begin{tabular}{@{} l c *3c @{}}
    \toprule
    \textbf{First category} & \textbf{Second category} & \multicolumn{3}{c}{\textbf{Third category}}\\
    \cmidrule(lr{1em}){3-5}
    & &\textbf{Reconstruction loss} & \textbf{KLD} & \textbf{ELBO}\\    
    \midrule
    Reconstruction loss(1) & latent IF(4) &  Reconstruction latent IF(7) &  KLD latent IF(10) &  ELBO latent IF(13) \\
    KLD(2) & latent LOF(5) &  Reconstruction latent LOF(8) &  KLD latent LOF(11) &  ELBO latent LOF(14) \\
    ELBO(3) & latent OCSVM(6) &  Reconstruction latent OCSVM(9) &  KLD latent OCSVM(12) &  ELBO latent OCSVM(15)\\
    \bottomrule
  \end{tabular}%
}%
\caption{\textbf{Abbreviations for outlier scores.}There are a total of 15 outlier scores, each with its own reference number.\label{tab:outlier_score_abbrev}}
\end{table}

In the first category, there are three outlier scores: Reconstruction loss, KLD and ELBO (see Eq. \ref{eq:ELBO_outlier}). For Reconstruction loss and KLD loss, we actually use their negative as the outlier scores such that outliers have smaller values than inliers, consistent with ELBO. CVAE, trained on a dataset with no or few outliers, is more difficult to model an anomalous input during inference time, resulting in lower negative Reconstruction loss\cite{an2015variational}. Recent research compared the performance of Reconstruction loss, KLD loss, and ELBO for outlier detection and discovered that Reconstruction loss presents lower discriminative power than either KLD loss or ELBO \cite{zimmerer2019unsupervised}. 

In the second category, there are three outlier scores. We obtain a low-dimensional latent vector for each image (sampled from the approximated posterior distribution for that image). The latent vectors are then passed into three traditional unsupervised outlier detection techniques, namely Isolation Forest (IF), Local Outlier Factor (LOF), and One-Class Support Vector Machine (OCSVM). We briefly summarize these three methods now. 

The IF model \cite{liu2008isolation} presupposes that anomalies are infrequent and distinct in the feature space. It is based on decision trees, in which data points are partitioned in each step by a random splitting value of a randomly selected characteristic. Outliers, due to their distinct values away from the more frequent inliers, are discovered to be partitioned closer to the root of the tree (i.e. shorter average path length). LOF \cite{breunig2000lof} computes the density deviation of a given data point with respect to its local k-nearest neighbors, where $k$ is a predefined hyperparameter. Outliers are samples that have a significantly lower density compared to their surrounding neighbours. OCSVM \cite{scholkopf2001estimating} calculates a decision boundary, and any data points that fall beyond of that threshold are considered outliers.  We use the IsolationForest, LocalOutlierFactor and OneClassSVM functions in sklearn package. Except for the one indicating the outlier ratio, which was set at $0.5\%$ based on prior experience with the ADMANI dataset, the default hyperparameters were used.

In the third category, the Reconstruction loss, KLD loss, and ELBO are each added as an extra dimension to the latent vector. The newly created vectors are then fed into IF, LOF, and OCSVM, resulting in $3 \times 3 = 9$ outlier scores. The reasoning for this is that the generative loss and the latent vector may provide complementary information for outlier detection \cite{angiulli2020improving}. 

\subsection{Receiver operating characteristic curve and precision-recall curve}
\label{sec:ROC_PRC}
In this section we discuss the evaluation metrics for outlier detection performance of various outlier scores. Based on the confusion matrix, a contingency table that show the number of correct predictions and incorrect predictions per class (see Table \ref{tab:contingency_tab}), the definitions for true positive rate (TPR), false positive rate (FPR), precision, recall and F1-score are shown below.\\

\begin{enumerate}
    \item $\textrm{TPR} = \textrm{Recall} = \frac{\textrm{true positives}}{\textrm{true positives} \textrm{ + } \textrm{false negatives}}$;
    \item $\textrm{FPR} = \frac{\textrm{false positives}}{\textrm{false positives} \textrm{ + } \textrm{true negatives}}$;
    \item $\textrm{Precision} = \frac{\textrm{true positives}}{\textrm{true positives} \textrm{ + } \textrm{false positives}}$;
    \item $\textrm{F1-score} = \frac{2 \times \textrm{precision} \times \textrm{recall}}{\textrm{precision} \textrm{ + } \textrm{recall}}$.
\end{enumerate}

\begin{table}[h!]
\resizebox{0.8\textwidth}{!}{%
  \begin{tabular}{@{} l c l c l @{}}
    \toprule
    & & \textbf{True outliers} & & \textbf{True inliers} \\
    \midrule
    \textbf{Predicted Outliers} & & Number of true positives & & Number of false positives \\
    \midrule
    \textbf{Predicted inliers} & & Number of false negatives & & Number of true negatives \\
    \bottomrule
  \end{tabular}%
}%
\caption{\textbf{Contingency table between true and predicted outlier labels.}\label{tab:contingency_tab}.}
\end{table}

The receiver operating characteristic (ROC) curve is a graphical representation of a binary classifier's trade-off between TPR and FPR. Because the ratio of outliers to inliers is so unbalanced, the ROC curve can provide an excessively optimistic impression of the classifier's effectiveness \cite{davis2006relationship}. When there is a substantial skew in the class distribution, the precision-recall curve is a more informative option. The area under the ROC curve is referred to as AUROC, while the area under the precision-recall curve is referred to as AUPRC. Higher values for both AUROC and AUPRC indicate greater performance. The baseline for AUROC is 0.5 or random guessing, while for AUPRC, it is the proportion of true outliers in the data set, which is approximately 0.005. Note that the baseline for AUPRC of each outlier type is also set to be approximately 0.005 since we get a random sample of inliers (i.e. mammogram without any technical artefacts) to make the ratio of one particular outlier type remain 0.005. The maximum of a list of precision/recall/F1-score given a list of thresholds are also provided. 
 
\subsection{Preprocessing steps for the ADMANI dataset}
\label{sec:preprocess}
Mammograms with breast implants differ significantly from other scans and are regarded as outliers. Because we already know if a breast image contains implants or not, we removed them from the ADMANI dataset. We then removed images where implants can be inferred based on the image manufacturer's algorithm attribute, leaving just images with the most commonly used methods. Following this sequence of filtering procedures, the ADMANI dataset's 30,000 images were reduced to 29,248.

Because nuisance variables contribute to image dissimilarities, reducing outlier identification systems' effectiveness \cite{fernando2021deep}, we began by removing the text artefacts from each image in the remaining ADMANI dataset. In each image, a minimal bounding rectangle surrounding the breast was clipped off and padded to a height/width ratio of two. This ratio has been chosen to minimize the padding required after cropping for this particular dataset. To further eliminate unwanted factors, left and right breast scans were normalized by mirroring all right breast images(see Figure S2 in supplementary file).

\subsection{Using true outliers in the ADMANI dataset for reference}
\label{sec:Reference_TrueOutliers}
To determine the reference true outliers in the ADMANI dataset, we combined deep learning with a well-trained radiologist.and obtained two reference lists, the union of which were confirmed as the final reference. To get the first reference list, 

\begin{enumerate}
    \item Train ResNetCVAE using the best configuration (see Section \ref{sec:CVAE_varying_depths}) on the preprocessed ADMANI dataset (see Section \ref{sec:preprocess}).
    \item Choose 600 images with the lowest scores for each of the 15 outlier metrics and obtain a union of the selected images.
    \item Have a non-radiologist choose true outliers from the union \footnote{At this point, the goal is to incorporate as many outliers as possible, even if they may be false positives.} and remove the selected true outliers from the ADMANI dataset.
    \item Train ResNetCVAE again on the leftover ADMANI dataset and repeat steps 2-3.
    \item Collect any true outliers identified by the non-radiologist in the two iterations and refer them to a professional radiologist for evaluation.
\end{enumerate}

To get the second reference list, 

\begin{enumerate}
    \item Using the preprocessed ADMANI dataset (see Section \ref{sec:preprocess}), train ResNetCVAE with the best setup (see Section \ref{sec:CVAE_varying_depths}).
    \item Choose the 200 images with the lowest scores for each of the 15 outlier metrics and get a union of all of the selected images.
    \item Remove the union of all selected images from the ADMANI dataset and train ResNetCVAE again on the leftover dataset.
    \item Repeat step 2).
    \item Combine the two removed unions and determine potential true outliers by a non-radiologist, which is further confirmed by a professional radiologist.
\end{enumerate}

The union of the two reference lists contains 136 true outliers, and the outlier ratio in the ADMANI dataset is $136/29248=0.465\%$. 

\begin{figure}[!ht]
  \begin{subfigure}[t]{0.22\textwidth}
  \includegraphics[width=0.9\textwidth]{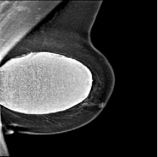}
  \end{subfigure}
  \vspace{0.5cm}
  \hfill
  \begin{subfigure}[t]{0.22\textwidth}  
  \includegraphics[width=0.9\textwidth]{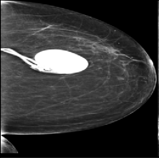}
  \end{subfigure}
  \hfill
  \begin{subfigure}[t]{0.22\textwidth}
  \includegraphics[width=0.9\textwidth]{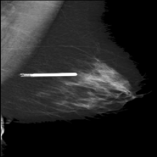}
  \end{subfigure}
  \hfill
  \begin{subfigure}[t]{0.22\textwidth}
  \includegraphics[width=0.9\textwidth]{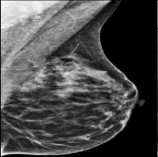}
  \end{subfigure}
  \vfill
  \begin{subfigure}[t]{0.22\textwidth}
  \includegraphics[width=0.9\textwidth]{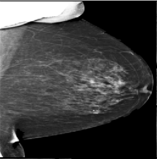}
  \end{subfigure}
  \hfill
  \begin{subfigure}[t]{0.22\textwidth}
  \includegraphics[width=0.9\textwidth]{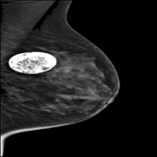}
  \end{subfigure}
  \hfill
  \begin{subfigure}[t]{0.22\textwidth}
  \includegraphics[width=0.9\textwidth]{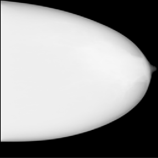}
  \end{subfigure}
  \hfill
  \begin{subfigure}[t]{0.22\textwidth}
  \includegraphics[width=0.9\textwidth]{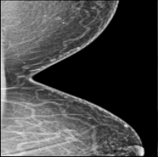}
  \end{subfigure}
  \caption{\textbf{Examples of true outlier subgroups in the ADMANI dataset.} From left to right and top to bottom, the outlier types are: implant, pacemaker, cardiac loop recorder, improper radiography, improper radiography, atypical lesions/calcification, incorrect exposure parameter, improper placement. Improper radiography is classified into two subtypes: those with heterogeneous pectoral muscle and the rest, respectively.\label{fig:representative_TrueOutliers}}    
\end{figure}

The 136 true outliers are classified into seven categories by the professional radiologist: implant, pacemaker, cardiac loop recorder, improper radiography, atypical lesion/calcification, incorrect exposure parameter and improper placement. It is worth noting that, despite the fact that mammograms with implants were eliminated during the preprocessing step (see Section \ref{sec:preprocess}), there were still mammograms with implants remaining since they were incorrectly categorised as having no implants. Figure \ref{fig:representative_TrueOutliers} depicts representative outliers for each category, and Table. \ref{tab:outlier_freq} displays the number and percentage of outliers in each category given that that the total number of true outliers is 136. 

\begin{table}[h!]
\resizebox{0.6\textwidth}{!}{%
\begin{tabular}{lcc}
\hline
Outlier type & Quantity & Percentage \\ \hline
Implants & 36 & 0.265\\
Pacemaker & 25 & 0.184\\
Cardiac loop recorder & 6 & 0.044\\
Improper radiography subtype 1 & 38 & 0.279\\
Improper radiography subtype 2 & 18 & 0.132\\
Atypical lesions/calcifications & 9 &  0.066\\ 
Incorrect exposure parameter & 2 & 0.0147\\ 
Improper placement & 2 & 0.015\\ \hline
\end{tabular}
}%
\captionof{table}{\textbf{Frequency table for all the outlier subgroups.} Quantity and percentage of images for each outlier category.}
\label{tab:outlier_freq}
\end{table}

\subsection{Hand-crafted outlier detection methods \label{sec:erosion_muscle}}
\paragraph{Erosion} We designed hand-crafted techniques in 5 phases to discover outliers that produce bright regions in the image, such as pacemakers, cardiac loop recorders, unusual lesions, and implants: 1) preprocess images (see Section \ref{sec:preprocess}), 2) threshold all images by a predetermined value to produce binary counterparts, 3) erode binary images to remove scattered signals, 4) for each image, add the sums of all pixels, 5) Sort the image pixel sums in descending order and choose the top $1\%$, $2\%$ and $5\%$ images. 

In total 16 configurations of hyperparameters were searched, i.e. 4 thresholds (180, 200, 220 and 240) cross 2 kernel sizes (5, 10) cross 2 iteration numbers (5, 10) for erosion. Supplementary Table S1 displays the mean recall rate for both training and testing set if $1\%$, $2\%$ and $5\%$ images are picked. The training and testing set are split once at a ratio of 0.6:0.4. The mean recall rate is averaged over 20 bootstraps of either training or testing set.  

\paragraph{Pectoral Muscle Analysis} To discover outliers with highly heterogeneous pectoral muscle regions \footnote{The pectoral muscle can only be observed in the mediolateral oblique view position (MLO).}: 1) preprocess images (see Section \ref{sec:preprocess}), 2) extract the pectoral muscle region, 3) count major lines in the muscle region, with greater line counts indicating outliers with more heterogeneous pectoral muscle regions. 

To extract the pectoral muscle region in step 2, we specifically followed the procedures in \cite{beeravolu2021preprocessing}: a) Resize the preprocessed images (see Section \ref{sec:preprocess}) to $256 \times 256$, b) Use canny edge detection to obtain strong edges in the image. Canny edge detection employs a Gaussian filter with $\sigma=5$ to remove noise, $3 \times 3$ Sobel kernels to calculate intensity gradient, non-maximum suppression to suppress pixels with no maximum gradient along the edge direction, and hysteresis thresholding to pick pixels with significant intensity gradient (pixels on strong edges) \footnote{Use the default lower and higher hysteresis thresholds for the skimage.feature.canny function, which are respectively $10\%$ and $20\%$ of the maximum gradient intensity for each image.}, c) apply the Hough line transformation to the canny edges and identify lines that are parameterized in angles and distances with respect to the center of the image. The lines with distances in the range $lower distance, upper distance]$ and angles in the range $(10^{\circ}, 70^{\circ})$ were recorded, and the line with the shortest distance was chosen as the muscle border and guides the removal of the pectoral muscle. We tried two choices for the lower distance, 5 and 20, and one option for the upper distance, 182.

In step 3, we used canny edge detection and the Hough line transform to count the principal lines in the pectoral muscle region. The aperture size for the Sobel operator for canny edge detection was 3; we tested 160 and 170 for lower thresholds, 180 and 220 for upper thresholds in hysteresis thresholding. We utilised the OpenCV library for the Hough line transformation, with $\rho$ set to 1, $\theta$ set to $\pi / 180$ \cite{beeravolu2021preprocessing}, and selection thresholds $40, 50, 60$. The images were then ranked in descending order by the number of lines in their pectoral muscle regions, with the top $1\%$, $2\%$, and $5\%$ images chosen as probable outliers. Images with line numbers greater than 8 were not considered in the ranking since the cut regions for these images from step 2 include not only pectoral muscle but also breast.

In total 24 configurations of hyperparameters were searched, i.e. 2 lower distances for muscle boundary selection in step 2 cross 2 lower thresholds cross 2 upper thresholds cross 3 selection thresholds in step 3. The mean recall rate is shown in Supplementary Table S2 if the top $1\%$, $2\%$ and $5\%$ images are chosen when they are ranked in descending order by number of lines in their pectoral muscle regions. The training and testing set are split once at the ratio of 0.6 to 0.4. The mean is calculated over 20 bootstraps for either the training or testing datasets.

\section{Experiments}
\subsection{Outlier detection using CVAE \label{sec:result_CVAE}}
\begin{figure}[h!]
    \centering
    \includegraphics[scale=0.35]{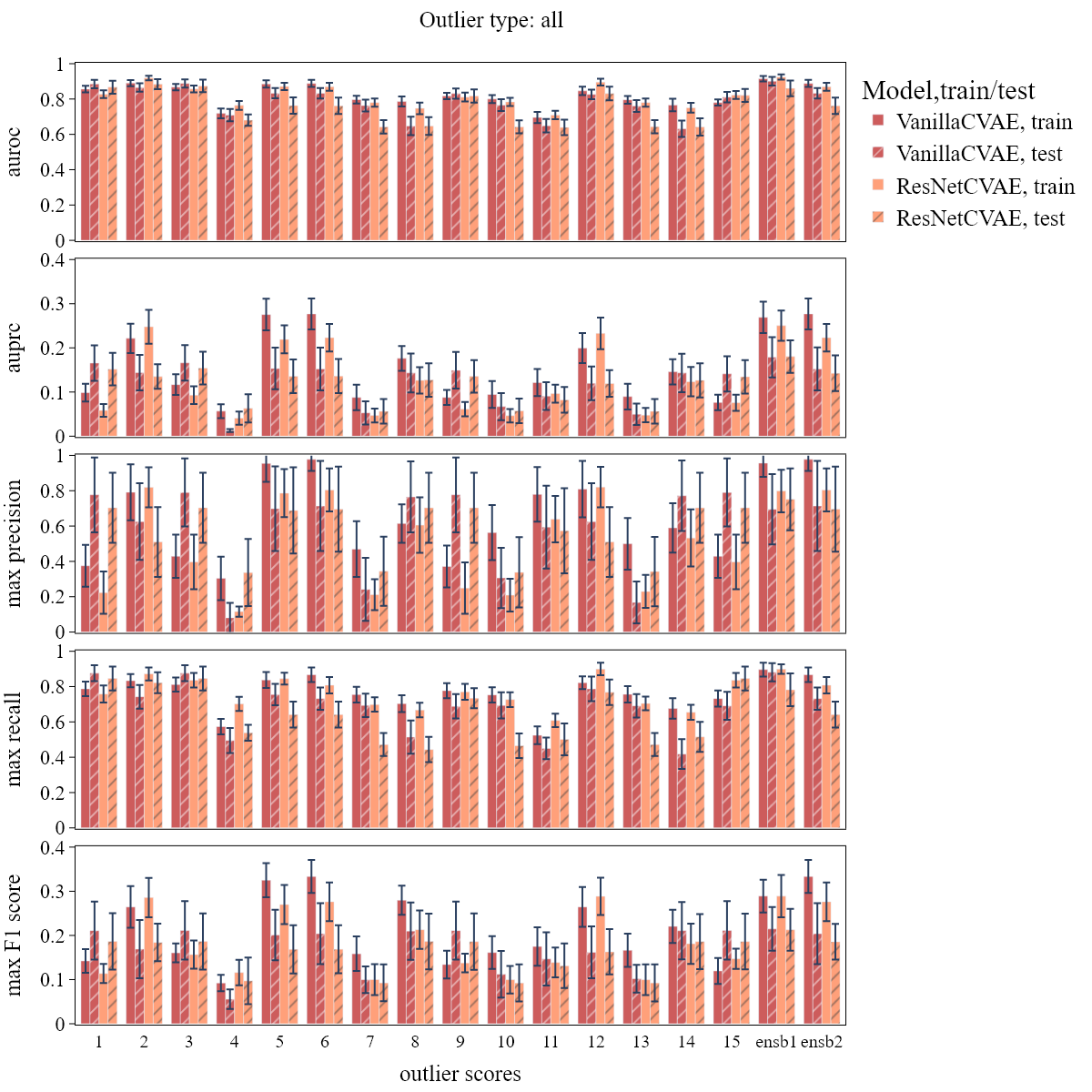}
    \caption{\textbf{Outlier detection performance for all 15 outlier scores.} For both VanillaCVAE and ResNetCVAE, the performance of all 15 outlier scores to detect outliers in the ADMANI dataset is shown in metrics: AUROC, AUPRC, max precision, max recall and max F1 score. To show the details of AUPRC and max F1 score for each outlier score, the y-axis has been reduced to the range [0, 0.4]. The number and name for each outlier score can be referred to Table. \ref{tab:outlier_score_abbrev}. There is not statistically significant difference between VanillaCVAE and ResNetCVAE. Although latent OCSVM(number 6) is the outlier scores with top AUPRC, overall speaking, the AUPRC is not high enough.}
    \label{fig:TrueOutliers_boxplot}
\end{figure}

We used CVAE to find technical outliers in the preprocessed ADMANI dataset (Section \ref{sec:preprocess}). Figure \ref{fig:TrueOutliers_boxplot} depicts the performance of 15 outlier scores for both training and testing sets using two distinct CVAE architectures (VanillaCVAE and ResNetCVAE, see Section \ref{sec:CVAE_varying_depths}). Both neural networks were trained only once, but the metrics (AUROC/AUPRC/Max Precision/Max Recall/Max F1 Score, see Section \ref{sec:ROC_PRC}) were derived by averaging over 20 bootstraps for both the training and testing sets. To ensure consistent outlier composition (outlier types and ratios), we sampled with replacement inliers and each outlier type separately for both training and testing set before combining them. The maximum precision/recall/F1 score is the highest value of a list of mean precision/recall/F1 scores across a range of threshold ratios, which includes 0.06\% to 0.2\% with interval 0.02\%, 0.2\% to 0.8\% with interval 0.1\%, 0.8\% to 12\% with interval 0.2\%, and 12\% to 30\% with interval 2\%.

In Figure \ref{fig:TrueOutliers_boxplot}, VanillaCVAE and ResNetCVAE do not detect outliers statistically differently for any of the outlier scores. Since ResNetCVAE is more complex and takes longer to train, we employed VanillaCVAE in the following study. Although the AUROC is relatively high, the maximum AUPRC across the 15 outlier scores in the test set by VanillaCVAE is 0.166, suggesting the need for outlier detection performance enhancement. We begin by investigating the possible causes of low AUPRC and then assess the performance of each outlier type.

\begin{figure}
    \centering
    \includegraphics[scale=0.3]{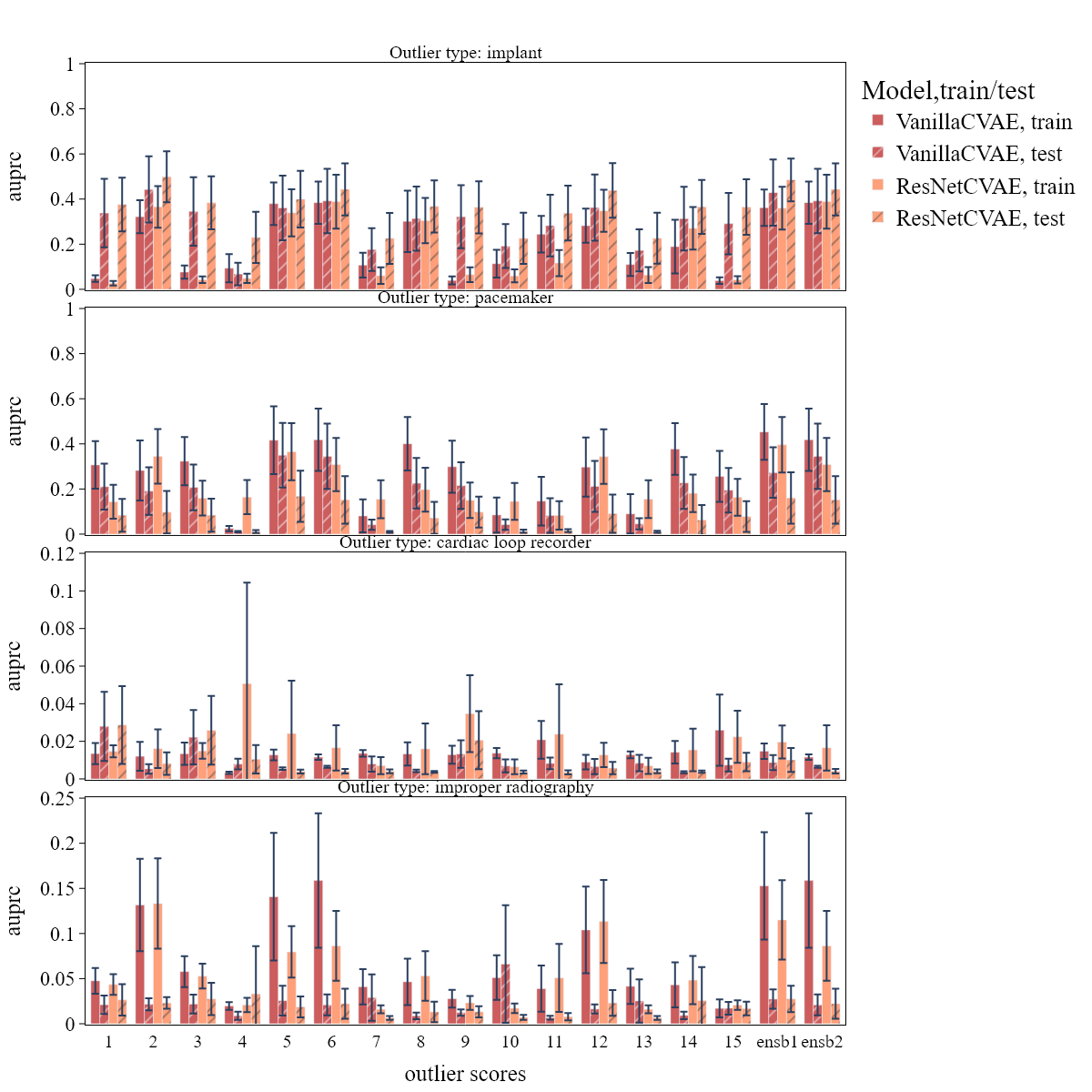}
    \caption{\textbf{AUPRC of outlier scores for each specific outlier category.} For each outlier category, the AUPRC across all 15 outlier scores is displayed. The number on the x-axis corresponds to outlier scores in Table \ref{tab:outlier_score_abbrev}. The ensb1 and ensb2 values are the average and minimum ensemble of three best performing outlier scores: Reconstruction loss, KLD and latent OCSVM. To show the AUPRC details for each outlier score, the y-axis for the cardiac loop recorder and improper radiology has been reduced to the range [0, 0.12] and [0, 0.25]. Implants and pacemakers are discovered at a higher rate than cardiac loop recorder and improper radiography. Different outlier scores work differently in recognising each outlier type. In outlier detection, the average ensemble (ensb1) is comparable to or better than the minimal ensemble (ensb2).}
    \label{fig:performance_specific_outlier}
\end{figure}

The AUPRC of the 15 outlier scores for each outlier type is shown in Figure \ref{fig:performance_specific_outlier} and supplementary Figure S3. As expected, VanillaCVAE performs differently in different outlier types: it has higher maximum AUPRC (across all 15 outlier scores in the test set) for outliers with implants (maximum AUPRC: 0.442), pacemaker (maximum AUPRC: 0.349) (Figure \ref{fig:performance_specific_outlier}), incorrect exposure parameter (maximum AUPRC: 1.0) and improper placement (maximum AUPRC: 0.687) (supplementary Figure S3); however lower maximum AUPRC in detecting outliers with cardiac loop recorder (maximum AUPRC: 0.028), improper radiography (maximum AUPRC: 0.066) (Figure \ref{fig:performance_specific_outlier}) and atypical lesions/calcification (maximum AUPRC: 0.155)(supplementary Figure S3). 

For outlier types with high AUPRC, different outlier scores behave differently: latent OCSVM has greater AUPRC than reconstruction loss for outliers with implants and pacemakers (Figure \ref{fig:performance_specific_outlier}), whereas reconstruction loss is better for outliers with improper placement (supplementary Figure S3). KLD has similar performance as latent OCSVM for outliers with implants (Figure \ref{fig:performance_specific_outlier}) and improper placement (supplementary Figure S3), but similar performance as Reconstruction loss for outliers with pacemaker (Figure \ref{fig:performance_specific_outlier}), which suggests that KLD loss may contain extra information for outlier detection which is also consistent with the fact that KLD loss is mathematically different from Reconstruction loss and latent OCSVM (see Section \ref{sec:outlier_scores}). In this study, for other outlier scores (numbers 7 to 15), adding Reconstruction loss, KLD loss, and ELBO as extra dimensions to the latent vector that are then applied to traditional outlier detection methods (e.g. IF, LOF, OCSVM, see Section \ref{sec:outlier_scores}) does not help to improve outlier detection performance.

Given that no outlier score excels for all outlier categories, we selected Reconstruction loss, KLD, and latent OCSVM and combined the three into a single indicator. The first ensemble takes the average of the three, whereas the second takes the smallest of the three. Because the three outlier scores are on different scales, they were first min-max normalised before being ensembled. Figures \ref{fig:performance_specific_outlier} and S3 demonstrate how the two ensembles (ensb1 and ensb2) perform for each outlier type. The average ensemble, in particular, is either comparable or better than the minimal ensemble at findings outliers, and will be utilised as the sole outlier indicator in the following study.

\subsection{Improve outlier detection performance using CVAE}
In the previous section, we examined CVAE's outlier detection ability on the ADMANI dataset. The benefit of CVAE is that it detects a variety of outlier types without prior expert knowledge and has satisfactory performance in some outlier kinds (Figures \ref{fig:performance_specific_outlier} and S3). Poorly detected outlier types can be easily discovered using CVAE's outlier type information and specific hand-crafted image processing approaches.

\begin{figure}[!ht]
  \begin{subfigure}[t]{0.3\textwidth}
  \caption{\label{fig:erosion_original}}
  \includegraphics[width=0.9\textwidth, left]{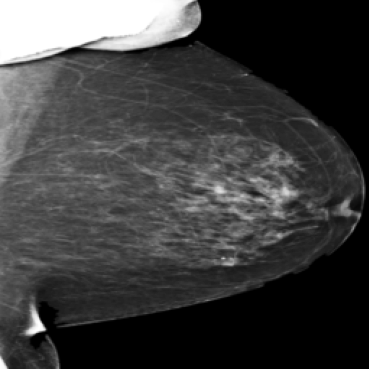}
  \end{subfigure}
  \hfill
  \begin{subfigure}[t]{0.3\textwidth}
  \caption{\label{fig:erosion_threshold}}
  \includegraphics[width=0.9\textwidth]{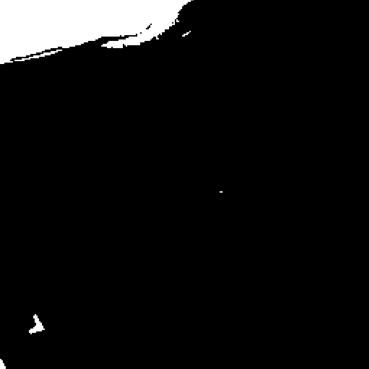}
  \end{subfigure}
  \hfill
  \begin{subfigure}[t]{0.3\textwidth}
  \caption{\label{fig:erosion_erosion}}
  \includegraphics[width=0.9\textwidth, right]{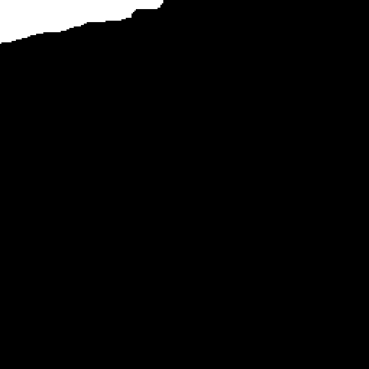}
  \end{subfigure}
  \vspace{0.5cm}
  \vfill
  \begin{subfigure}[t]{0.3\textwidth}  \caption{\label{fig:pectoral_muscle_original}}
  \includegraphics[width=0.9\textwidth, left]{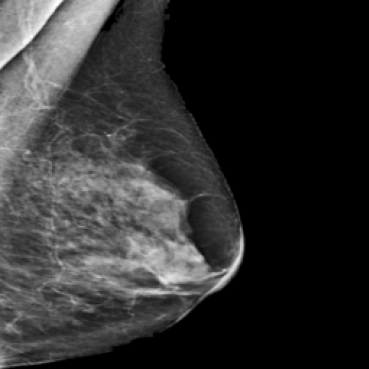}
  \end{subfigure}
  \hfill
  \begin{subfigure}[t]{0.3\textwidth}
  \caption{\label{fig:pectoral_muscle_cut}}
  \includegraphics[width=0.9\textwidth]{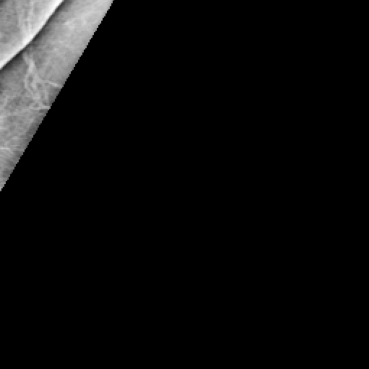}
  \end{subfigure}
  \hfill
  \begin{subfigure}[t]{0.3\textwidth}
  \caption{\label{fig:pectoral_muscle_canny}}
  \includegraphics[width=0.9\textwidth, right]{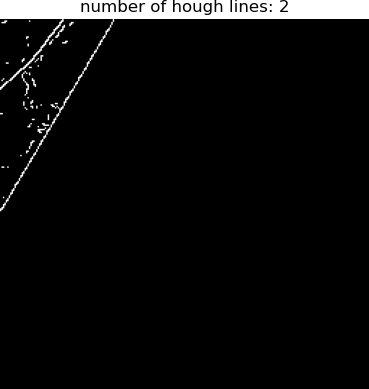}
  \end{subfigure}
  \caption{\textbf{Image processing techniques for outlier detection.} Panels A, B and C show the image erosion process for outlier detection. A fixed threshold transforms the original outlier in Panel A into the binary image in Panel B. Panel B' scattered signal gets eroded, as demonstrated in Panel C. Outliers present a large bright region in Panel C. Panel D, E, and F show how to perform pectoral muscle analysis for outlier detection. Panel E removes the pectoral muscle region of the original outlier in Panel D, and Panel F displays the major lines in the pectoral muscle region. A higher number denotes an anomaly. \label{fig:erosion}}
\end{figure}

CVAE, as previously observed, performs poorly in detecting outliers with cardiac loop recorder, improper radiography and atypical lesions/calcifications (Figures \ref{fig:performance_specific_outlier} and S3). The presence of bright regions (unrelated to cancer tissue) is a common feature of these poorly recognized outliers (Figure \ref{fig:representative_TrueOutliers}); as a result, it is natural to utilise the presence of such regions as a signal of outliers. The major steps are shown in Figure \ref{fig:erosion} where an outlier image due to improper radiography (Figure \ref{fig:erosion_original}) was first converted to a white-black binary image thresholded by some fixed pixel value (Figure \ref{fig:erosion_threshold}), and then eroded such that only the bright region remains (Figure \ref{fig:erosion_erosion}). The image is more likely to be an outlier if the remaining region is larger. Detailed methodology can be found in Section \ref{sec:erosion_muscle}.

There is a unique subgroup of improper radiography outliers with very diverse pectoral muscle region. Such an outlier is seen in Figure \ref{fig:pectoral_muscle_original}, and it is found by first cutting the pectoral muscle region out (Figure \ref{fig:pectoral_muscle_cut}) and then counting the major line number in the pectoral muscle region (Figure \ref{fig:pectoral_muscle_canny}). The greater the number, the greater the likelihood that the image is an outlier. Detailed information can be found in Section \ref{sec:erosion_muscle}.

Table \ref{tab:performance_VanillaCVAE_erosion_muscle} shows the true reference outlier number as well as the outlier detection performance by VanillaCVAE, erosion and pectoral muscle analysis for both the training and testing set. The results for the VanillaCVAE is the outlier recall rate if the top $1\%$, $2\%$, or $5\%$ images ranked in ascending order by the average ensemble outlier score (see Section \ref{sec:result_CVAE})) are selected; the results for erosion are the recall rate if \textbf{VanillaCVAE} is complemented by erosion where erosion selects the top $1\%$, $2\%$ or $5\%$ images ranked in descending order by pixel sums (see Section \ref{sec:erosion_muscle}); and the results for the pectoral muscle analysis are the recall rate if \textbf{VanillaCVAE and erosion} are complemented by pectoral muscle analysis where pectoral muscle analysis selects the top $1\%$, $2\%$ and $5\%$ images ranked in descending order by line numbers (see Section \ref{sec:erosion_muscle}). VanillaCVAE is trained only once and the train, valid and test split ratio is 0.6:0.1:0.4. All percentages are mean values across 20 bootstraps for both the training and testing dataset evaluated by the trained VanillaCVAE.

Erosion significantly increases the recall rate in addition to VanillaCVAE, while pectoral muscle analysis moderately boosts the recall rate in addition to VanillaCVAE and erosion. VanillaCVAE's recall rate in the test set is $0.31 \pm 0.06$ if the top $1\%$ images are chosen. When VanillaCVAE is complemented by erosion, the recall rate is $0.56 \pm 0.07$, with $0.25$ recall rate increase. When VanillaCVAE and erosion is complemented further by pectoral muscle analysis, the recall rate is $0.61 \pm 0.07$, with $0.05$ recall rate increase. If VanillaCVAE, erosion, and pectoral muscle analysis each select $5\%$ images, the recall rate rises to $83\%$.

\begin{table}
\resizebox{0.95\textwidth}{!}{%
  \begin{tabular}{@{} l *3c *3c *3c@{}}
    \toprule
    \multirow{2}{*}{\thead{\textbf{Reference}\\ \textbf{Number}}} &  \multicolumn{3}{c}{\textbf{1\%}}&\multicolumn{3}{c}{\textbf{2\%}} & \multicolumn{3}{c}{\textbf{5\%}}\\
    \cmidrule(lr{1em}){2-4}\cmidrule(lr{1em}){5-7}\cmidrule(lr{1em}){8-10}
    &{\textbf{VanillaCVAE}}&{\textbf{Erosion}}&{\textbf{Pectoral Muscle Analysis}}&{\textbf{VanillaCVAE}}&{\textbf{Erosion}}&{\textbf{Pectoral Muscle Analysis}}&{\textbf{VanillaCVAE}}&{\textbf{Erosion}}&{\textbf{Pectoral Muscle Analysis}}\\
\midrule
\multicolumn{10}{l}{\textbf{Train}}\\
\midrule 
87 & $0.35 \pm 0.04$ & $0.55 \pm 0.03$ & $0.55 \pm 0.03$ & $0.49 \pm 0.05$ & $0.7 \pm 0.05$ & $0.72 \pm 0.05$ & $0.63 \pm 0.06$ & $0.84 \pm 0.05$ & $0.86 \pm 0.05$ \\
\midrule
\multicolumn{10}{l}{\textbf{Test}}\\
\midrule
39 & $0.31 \pm 0.06$ & $0.56 \pm 0.07$ & $0.61 \pm 0.07$ &  $0.37 \pm 0.07$ & $0.64 \pm 0.07$ & $0.69 \pm 0.07$ & $0.54 \pm 0.06$ & $0.75 \pm 0.06$ & $0.83 \pm 0.06$\\
\bottomrule
\end{tabular}
}%
\caption{\textbf{Outlier detection performance using VanillaCVAE, erosion, and pectoral muscle analysis.} In this table, for both training and testing dataset, we show the reference true outlier number, the recall rate by VanillaCVAE if the top $1\%$, $2\%$ or $5\%$ images ranked in ascending order by average ensembled outlier score are selected, the recall rate if VanillaCVAE is complemented by erosion where erosion selects the top $1\%$, $2\%$ or $5\%$ images ranked in descending order by pixel sums (see Section \ref{sec:erosion_muscle}), and the recall rate if VanillaCVAE and erosion is complemented by pectoral muscle analysis where the pectoral muscle analysis selects the top $1\%$, $2\%$ or $5\%$ images ranked in descending order by line numbers (see Section \ref{sec:erosion_muscle}).\label{tab:performance_VanillaCVAE_erosion_muscle}}
\end{table}

\section{Conclusions and future work}
As a first step, we use deep learning and traditional image processing approaches to detect technical outliers in the ADMANI dataset, which contains breast cancer mammograms. Because there are no outlier labels available, we utilize CVAE, an unsupervised deep learning method, to scan the dataset first and learn about the types of outliers. CVAE performs differently for different outlier types, and classic image processing techniques like erosion and pectoral muscle analysis supplement CVAE in recognizing specific outlier types that it fails to detect. We recall $56\%$ of outliers if VanillaCVAE, erosion and pectoral muscle analysis each select $1\%$ images of the test set, and $81\%$ if each select $5\%$ images.

The reference true outliers in this research are obtained by deep learning screening with relaxed criteria first, then selected by a unprofessional research staff and finally confirmed by a professional radiologist. It only includes outlier types with obvious technical artefacts such as implanted medical devices but excludes those easily ignored by either deep learning or the unprofessional research staff such as antiperspirant, dust artefacts due to dust on compression paddle etc. Participation of professional radiologists to examine each image in the whole dataset will provide a more inclusive reference list. Professional expert knowledge and traditional image processing techniques may have potential to detect technical outliers that are not detected in this research.

For outlier types that could be discovered by deep learning, increasing the detection performance is a remaining task in the future. One direction is to try novel and efficient outlier score for VAE, such as likelihood regret which measures the log likelihood improvement of VAE configuration that maximizes the likelihood of an individual sample over the configuration that maximizes the likelihood of the whole dataset
\cite{xiao2020likelihood}. A second direction is to obtain better latent representations using algorithms such as VQ-VAE (see sec. \ref{sec:introduction}), or obtaining latent representations optimizing the objective of outlier detection directly such as OC-NN (see sec. \ref{sec:introduction}) or the combination of VQ-VAE and OC-NN. With the accumulation of more outlier labels determined by unsupervised learning, combined with data augmentation to increase outlier variation and proportion, the technical outlier detection problem in the ADMANI dataset could gradually change into supervised learning.



\bibliography{reference}

\begin{thebibliography}{10}
\expandafter\ifx\csname url\endcsname\relax
  \def\url#1{\texttt{#1}}\fi
\expandafter\ifx\csname urlprefix\endcsname\relax\def\urlprefix{URL }\fi
\expandafter\ifx\csname href\endcsname\relax
  \def\href#1#2{#2} \def\path#1{#1}\fi

\bibitem{perry2008european}
N.~Perry, M.~Broeders, C.~de~Wolf, S.~T{\"o}rnberg, R.~Holland, L.~von Karsa,
  European guidelines for quality assurance in breast cancer screening and
  diagnosis. -summary document, Oncology in Clinical Practice 4~(2) (2008)
  74--86.

\bibitem{van2015right}
C.~van Landsveld-Verhoeven, The right focus: manual on mammography positioning
  technique, LRCB, 2015.

\bibitem{geiser2011challenges}
W.~R. Geiser, T.~M. Haygood, L.~Santiago, T.~Stephens, D.~Thames, G.~J.
  Whitman, Challenges in mammography: part 1, artifacts in digital mammography,
  American Journal of Roentgenology 197~(6) (2011) W1023--W1030.

\bibitem{ayyala2008digital}
R.~S. Ayyala, M.~Chorlton, R.~H. Behrman, P.~J. Kornguth, P.~J. Slanetz,
  Digital mammographic artifacts on full-field systems: what are they and how
  do i fix them?, Radiographics 28~(7) (2008) 1999--2008.

\bibitem{paap2016mammography}
E.~Paap, M.~Witjes, C.~van Landsveld-Verhoeven, R.~M. Pijnappel, A.~H. Maas,
  M.~J. Broeders, Mammography in females with an implanted medical device:
  impact on image quality, pain and anxiety, The British Journal of Radiology
  89~(1066) (2016) 20160142.

\bibitem{hawkins1980identification}
D.~M. Hawkins, Identification of outliers, Vol.~11, Springer, 1980.

\bibitem{chalapathy2019deep}
R.~Chalapathy, S.~Chawla, Deep learning for anomaly detection: A survey, arXiv
  preprint arXiv:1901.03407 (2019).

\bibitem{fernando2020neural}
T.~Fernando, S.~Denman, D.~Ahmedt-Aristizabal, S.~Sridharan, K.~R. Laurens,
  P.~Johnston, C.~Fookes, Neural memory plasticity for medical anomaly
  detection, Neural Networks 127 (2020) 67--81.

\bibitem{fernando2021deep}
T.~Fernando, H.~Gammulle, S.~Denman, S.~Sridharan, C.~Fookes, Deep learning for
  medical anomaly detection--a survey, ACM Computing Surveys (CSUR) 54~(7)
  (2021) 1--37.

\bibitem{koufakou2009scalable}
A.~Koufakou, Scalable and efficient outlier detection in large distributed data
  sets with mixed-type attributes, University of Central Florida, 2009.

\bibitem{thudumu2020comprehensive}
S.~Thudumu, P.~Branch, J.~Jin, J.~J. Singh, A comprehensive survey of anomaly
  detection techniques for high dimensional big data, Journal of Big Data 7~(1)
  (2020) 1--30.

\bibitem{radovanovic2010hubs}
M.~Radovanovic, A.~Nanopoulos, M.~Ivanovic, Hubs in space: Popular nearest
  neighbors in high-dimensional data, Journal of Machine Learning Research
  11~(sept) (2010) 2487--2531.

\bibitem{radovanovic2014reverse}
M.~Radovanovi{\'c}, A.~Nanopoulos, M.~Ivanovi{\'c}, Reverse nearest neighbors
  in unsupervised distance-based outlier detection, IEEE transactions on
  knowledge and data engineering 27~(5) (2014) 1369--1382.

\bibitem{frazer2022admani}
H.~M. Frazer, J.~S. Tang, M.~S. Elliott, K.~M. Kunicki, B.~Hill, R.~Karthik,
  C.~F. Kwok, C.~A. Pe{\~n}a-Solorzano, Y.~Chen, C.~Wang, et~al., Admani:
  Annotated digital mammograms and associated non-image datasets, Radiology:
  Artificial Intelligence 5~(2) (2022) e220072.

\bibitem{lu2020universal}
Y.~Lu, J.~Lu, A universal approximation theorem of deep neural networks for
  expressing probability distributions, Advances in neural information
  processing systems 33 (2020) 3094--3105.

\bibitem{hawkins2002outlier}
S.~Hawkins, H.~He, G.~Williams, R.~Baxter, Outlier detection using replicator
  neural networks, in: International Conference on Data Warehousing and
  Knowledge Discovery, Springer, 2002, pp. 170--180.

\bibitem{tagawa2015structured}
T.~Tagawa, Y.~Tadokoro, T.~Yairi, Structured denoising autoencoder for fault
  detection and analysis, in: Asian conference on machine learning, PMLR, 2015,
  pp. 96--111.

\bibitem{chalapathy2017robust}
R.~Chalapathy, A.~K. Menon, S.~Chawla, Robust, deep and inductive anomaly
  detection, in: Joint European Conference on Machine Learning and Knowledge
  Discovery in Databases, Springer, 2017, pp. 36--51.

\bibitem{chen2018autoencoder}
Z.~Chen, C.~K. Yeo, B.~S. Lee, C.~T. Lau, Autoencoder-based network anomaly
  detection, in: 2018 Wireless telecommunications symposium (WTS), IEEE, 2018,
  pp. 1--5.

\bibitem{an2015variational}
J.~An, S.~Cho, Variational autoencoder based anomaly detection using
  reconstruction probability, Special Lecture on IE 2~(1) (2015) 1--18.

\bibitem{lu2018anomaly}
Y.~Lu, P.~Xu, Anomaly detection for skin disease images using variational
  autoencoder, arXiv preprint arXiv:1807.01349 (2018).

\bibitem{zimmerer2018context}
D.~Zimmerer, S.~A. Kohl, J.~Petersen, F.~Isensee, K.~H. Maier-Hein,
  Context-encoding variational autoencoder for unsupervised anomaly detection,
  arXiv preprint arXiv:1812.05941 (2018).

\bibitem{matias2021robust}
P.~Matias, D.~Folgado, H.~Gamboa, A.~V. Carreiro, Robust anomaly detection in
  time series through variational autoencoders and a local similarity score.,
  in: Biosignals, 2021, pp. 91--102.

\bibitem{xu2018unsupervised}
H.~Xu, W.~Chen, N.~Zhao, Z.~Li, J.~Bu, Z.~Li, Y.~Liu, Y.~Zhao, D.~Pei, Y.~Feng,
  et~al., Unsupervised anomaly detection via variational auto-encoder for
  seasonal kpis in web applications, in: Proceedings of the 2018 world wide web
  conference, 2018, pp. 187--196.

\bibitem{van2017neural}
A.~Van Den~Oord, O.~Vinyals, et~al., Neural discrete representation learning,
  Advances in neural information processing systems 30 (2017).

\bibitem{marimont2021anomaly}
S.~N. Marimont, G.~Tarroni, Anomaly detection through latent space restoration
  using vector quantized variational autoencoders, in: 2021 IEEE 18th
  International Symposium on Biomedical Imaging (ISBI), IEEE, 2021, pp.
  1764--1767.

\bibitem{xia2022gan}
X.~Xia, X.~Pan, N.~Li, X.~He, L.~Ma, X.~Zhang, N.~Ding, Gan-based anomaly
  detection: a review, Neurocomputing (2022).

\bibitem{li2018anomaly}
D.~Li, D.~Chen, J.~Goh, S.-k. Ng, Anomaly detection with generative adversarial
  networks for multivariate time series, arXiv preprint arXiv:1809.04758
  (2018).

\bibitem{donahue2016adversarial}
J.~Donahue, P.~Kr{\"a}henb{\"u}hl, T.~Darrell, Adversarial feature learning,
  arXiv preprint arXiv:1605.09782 (2016).

\bibitem{schlegl2017unsupervised}
T.~Schlegl, P.~Seeb{\"o}ck, S.~M. Waldstein, U.~Schmidt-Erfurth, G.~Langs,
  Unsupervised anomaly detection with generative adversarial networks to guide
  marker discovery, in: Information Processing in Medical Imaging: 25th
  International Conference, IPMI 2017, Boone, NC, USA, June 25-30, 2017,
  Proceedings, Springer, 2017, pp. 146--157.

\bibitem{chalapathy2018anomaly}
R.~Chalapathy, A.~K. Menon, S.~Chawla, Anomaly detection using one-class neural
  networks, arXiv preprint arXiv:1802.06360 (2018).

\bibitem{ruff2018deep}
L.~Ruff, R.~Vandermeulen, N.~Goernitz, L.~Deecke, S.~A. Siddiqui, A.~Binder,
  E.~M{\"u}ller, M.~Kloft, Deep one-class classification, in: International
  conference on machine learning, PMLR, 2018, pp. 4393--4402.

\bibitem{kingma2013auto}
D.~P. Kingma, M.~Welling, Auto-encoding variational bayes, arXiv preprint
  arXiv:1312.6114 (2013).

\bibitem{julianstastny}
{Github},
  \href{https://github.com/julianstastny/VAE-ResNet18-PyTorch}{Vae-resnet18-pytorch}
  (2019).
\newline\urlprefix\url{https://github.com/julianstastny/VAE-ResNet18-PyTorch}

\bibitem{he2016deep}
K.~He, X.~Zhang, S.~Ren, J.~Sun, Deep residual learning for image recognition,
  in: Proceedings of the IEEE conference on computer vision and pattern
  recognition, 2016, pp. 770--778.

\bibitem{zimmerer2019unsupervised}
D.~Zimmerer, F.~Isensee, J.~Petersen, S.~Kohl, K.~Maier-Hein, Unsupervised
  anomaly localization using variational auto-encoders, in: International
  Conference on Medical Image Computing and Computer-Assisted Intervention,
  Springer, 2019, pp. 289--297.

\bibitem{liu2008isolation}
F.~T. Liu, K.~M. Ting, Z.-H. Zhou, Isolation forest, in: 2008 eighth ieee
  international conference on data mining, IEEE, 2008, pp. 413--422.

\bibitem{breunig2000lof}
M.~M. Breunig, H.-P. Kriegel, R.~T. Ng, J.~Sander, Lof: identifying
  density-based local outliers, in: Proceedings of the 2000 ACM SIGMOD
  international conference on Management of data, 2000, pp. 93--104.

\bibitem{scholkopf2001estimating}
B.~Sch{\"o}lkopf, J.~C. Platt, J.~Shawe-Taylor, A.~J. Smola, R.~C. Williamson,
  Estimating the support of a high-dimensional distribution, Neural computation
  13~(7) (2001) 1443--1471.

\bibitem{angiulli2020improving}
F.~Angiulli, F.~Fassetti, L.~Ferragina, Improving deep unsupervised anomaly
  detection by exploiting vae latent space distribution, in: International
  Conference on Discovery Science, Springer, 2020, pp. 596--611.

\bibitem{davis2006relationship}
J.~Davis, M.~Goadrich, The relationship between precision-recall and roc
  curves, in: Proceedings of the 23rd international conference on Machine
  learning, 2006, pp. 233--240.

\bibitem{beeravolu2021preprocessing}
A.~R. Beeravolu, S.~Azam, M.~Jonkman, B.~Shanmugam, K.~Kannoorpatti, A.~Anwar,
  Preprocessing of breast cancer images to create datasets for deep-cnn, IEEE
  Access 9 (2021) 33438--33463.

\bibitem{xiao2020likelihood}
Z.~Xiao, Q.~Yan, Y.~Amit, Likelihood regret: An out-of-distribution detection
  score for variational auto-encoder, Advances in neural information processing
  systems 33 (2020) 20685--20696.

\end{thebibliography}
\bibliographystyle{elsarticle-num} 



%%\bibitem{}

%%\end{thebibliography}
\end{document}

\begin{figure}
\begin{tabular}{c @{\qquad} c @{\qquad} c}
    \includegraphics[width=.3\linewidth]{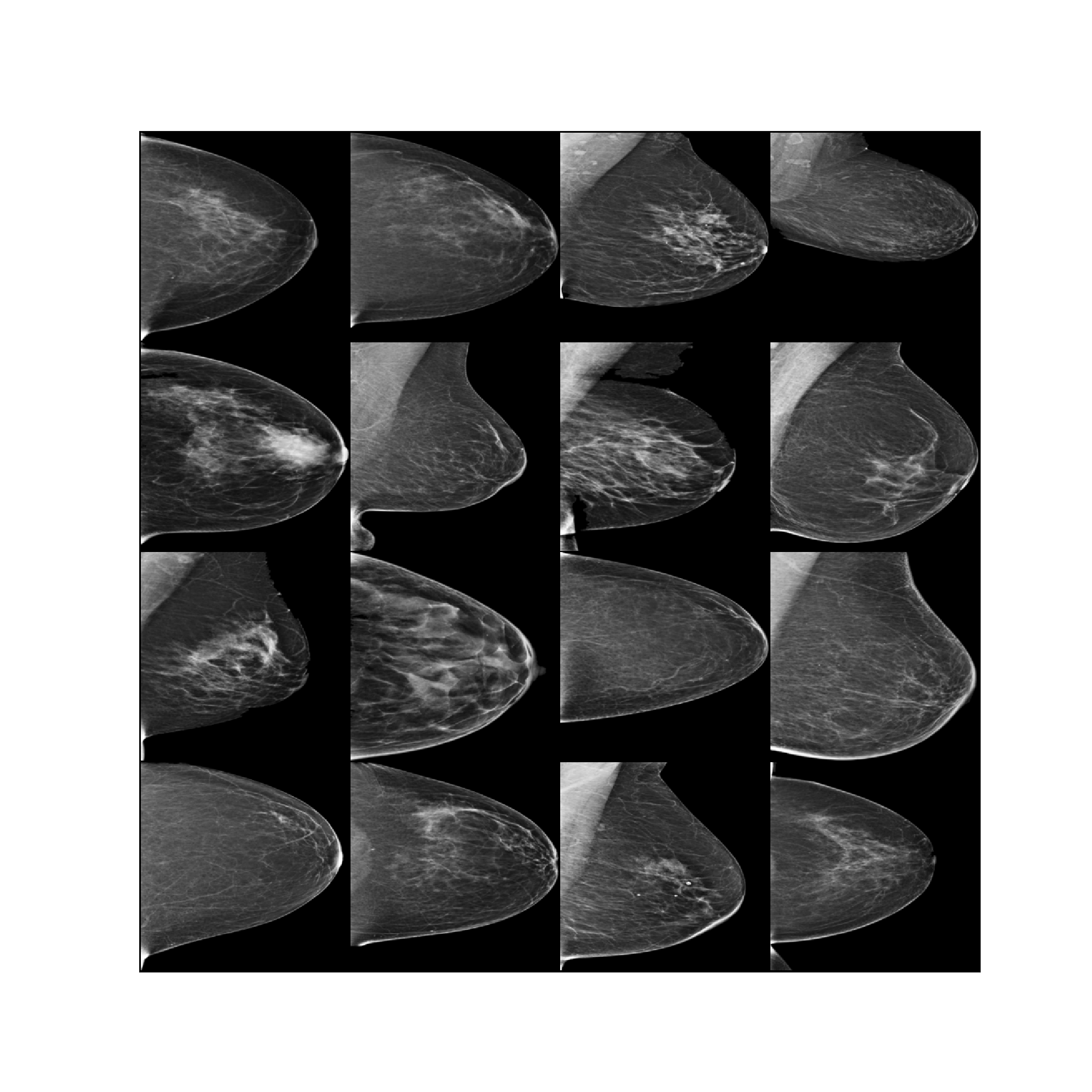} &
    \includegraphics[width=.2\linewidth]{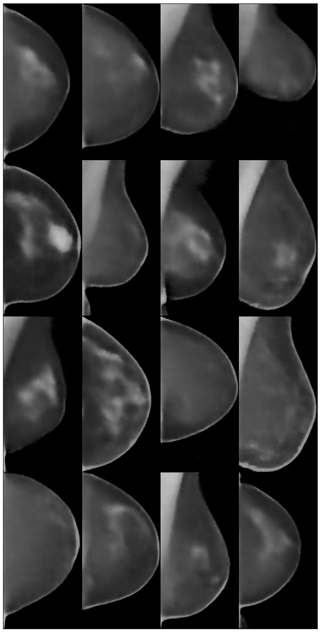} &
    \includegraphics[width=.3\linewidth]{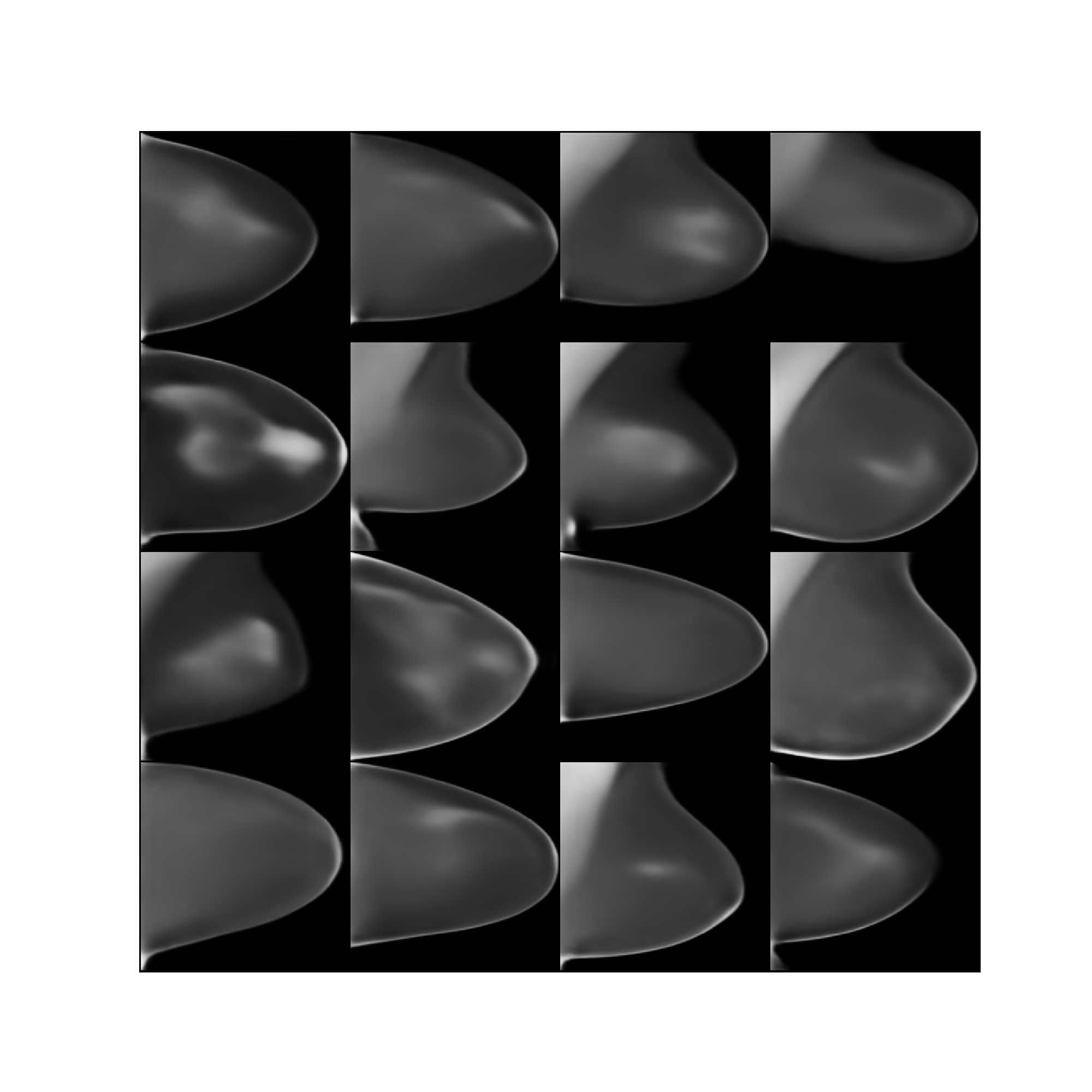} \\
    \small A & \small B & \small C
\end{tabular}
\caption{\textbf{The input mammogram and the generated output by VanillaCVAE and ResNetCVAE with their optimal configurations of hyperparameters.} Panel A shows the input mammogram in size $256 \times 256$; Panel B shows the generated output from VanillaCVAE, the best height $\times$ width for VanillaCVAE is $512 \times 256$; panel C shows the generated output from ResNetCVAE, the best height $\times$ width for ResNetCVAE is $256 \times 256$. The generated output image from VanillaCVAE is better than that from ResNetCVAE; the main reason is that ResNetCVAE is much deeper than VanillaCVAE and could hardly converge when image is resized in larger height or width (e.g. 512 versus 256 for height).\label{fig:best_config}}
\end{figure}

\begin{figure}
\begin{tabular}{c @{\qquad} c  c @{\qquad} c}
    \includegraphics[width=.48\linewidth]{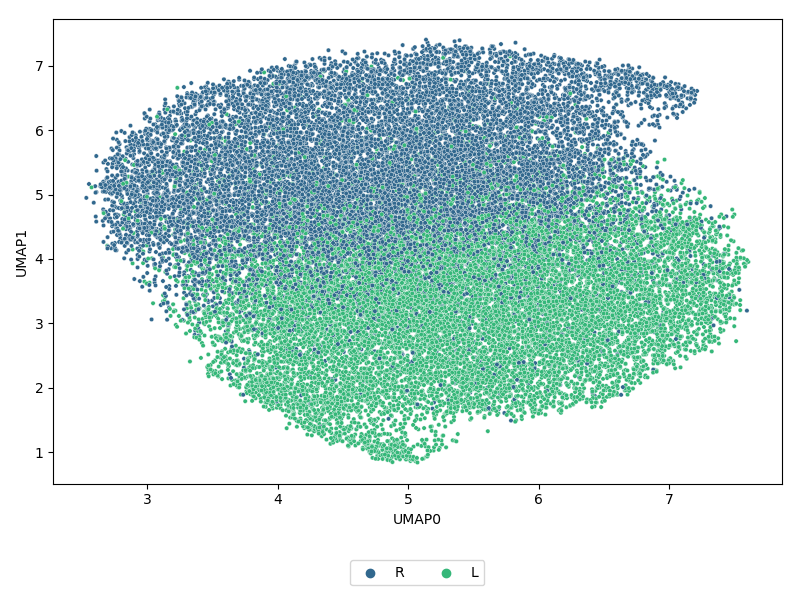} &
    \includegraphics[width=.48\linewidth]{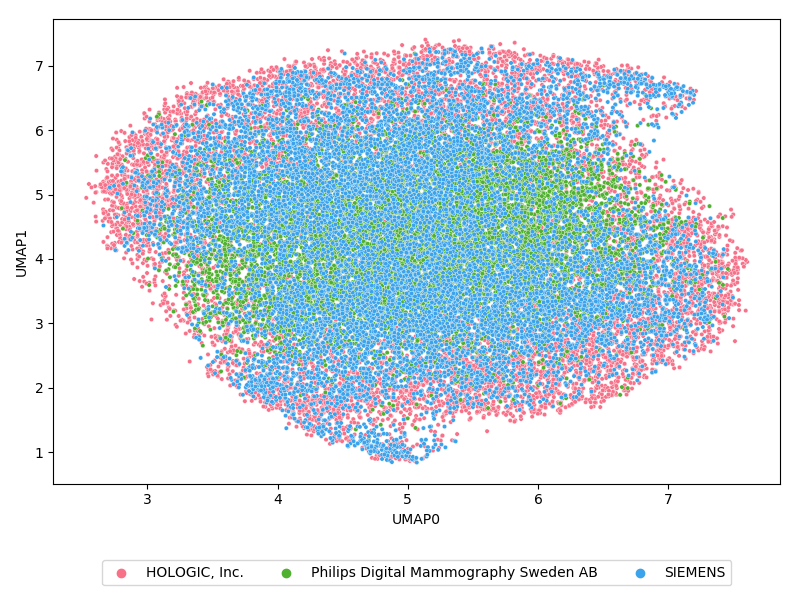} \\
    \small A \quad Image laterality, no mirror & \small B \quad Image manufacturer, no mirror \\ 
    \includegraphics[width=.48\linewidth]{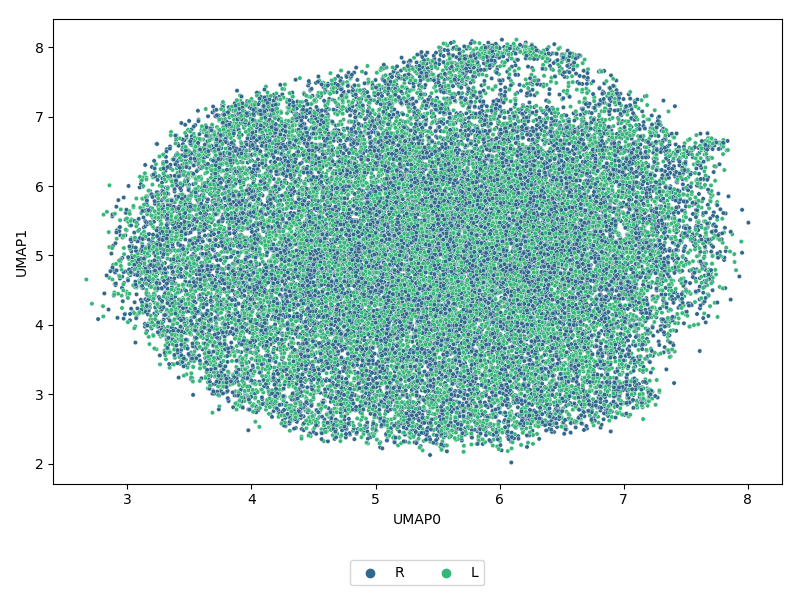} &
    \includegraphics[width=.48\linewidth]{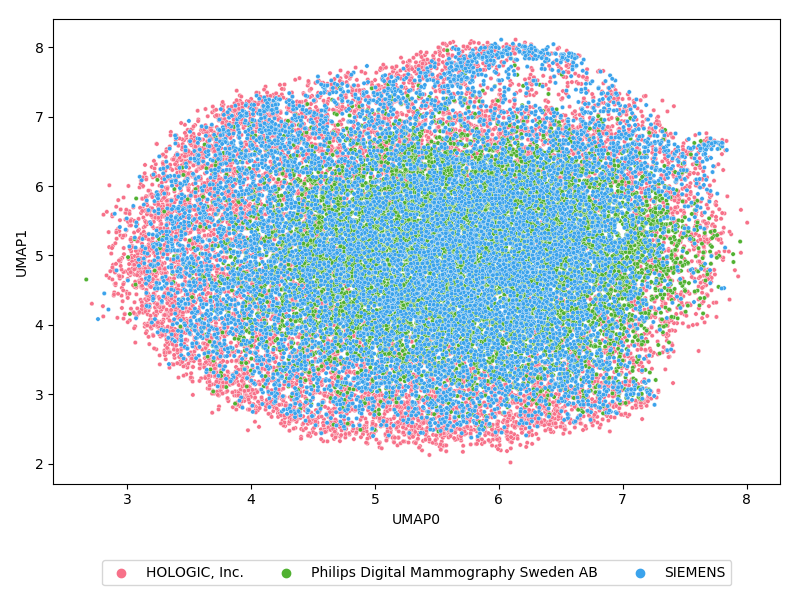}\\
    \small C \quad Image laterality, mirror & \small C \quad Image manufacturer, mirror
\end{tabular}
\caption{\textbf{Confounding effects in the ADMANI dataset.} There is large variability between left and right breasts (A). Mirroring right breast into left totally removes the variability due to image laterality (C). Although variance due to image manufacturer exist (B), it is rather trivial and there is little appropriate to do to remove it, we just leave it unchanged (D). \label{fig:confounding_effect}}
\end{figure}

\begin{figure}
\caption{\textbf{AUPRC of outlier scores on three specific outlier types.} We show AUPRC of all the 15 outlier scores for three specific outlier types: atypical lesions/calcification, incorrect exposure parameter and improper placement. AUPRC for both training and testing set from both VanillaCVAE and ResNetCVAE architectures are available. A caveat is that the two incorrect exposure parameter outliers are all in training set and the two improper placement outliers (Table. 4 in the main file) are all in the testing set. The number for the outlier scores can be referred in Table 2 in the main file. The y-axis for atypical lesion/calcification are shrunk to $[0, 0.5]$ to show auprc details for all outlier scores. Outliers with atypical lesion/calcification are hard to detect by CVAE. In contrast, outliers due to incorrect exposure parameter and improper placement are easy to detect by some outlier scores.  
\label{fig:auprc_loop_lesion_algorithm}}
    \centering
    \includegraphics[scale=0.3]{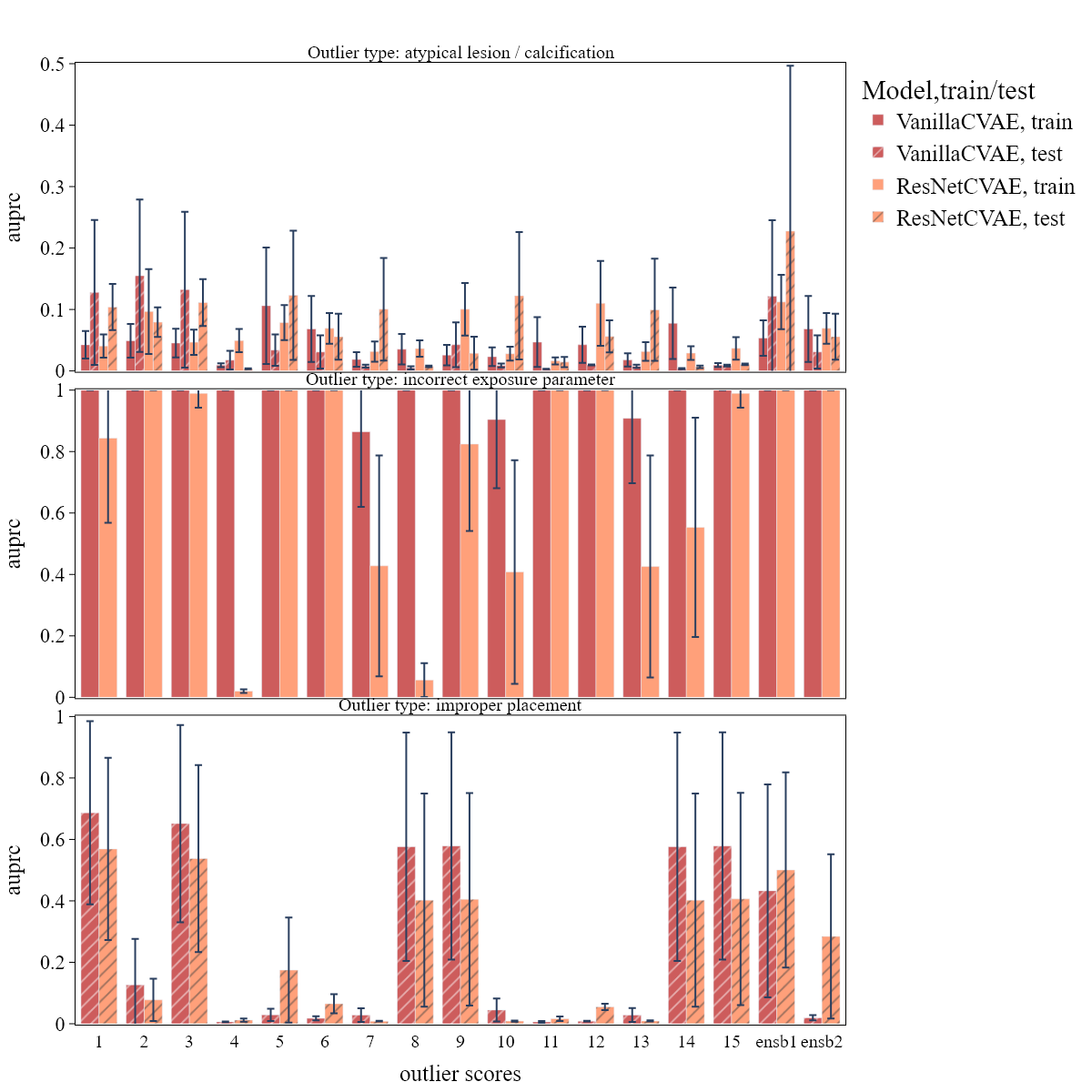}
\end{figure} 

\begin{table}
\resizebox{\textwidth}{!}{%
  \begin{tabular}{@{} lll *3c *3c@{}}
    \toprule
    \multirow{2}{*}{\thead{\textbf{Binary}\\ \textbf{Threshold}}} &\multirow{2}{*}{\thead{\textbf{Kernel}\\ \textbf{Size}}} & \multirow{2}{*}{\thead{\textbf{Iteration}\\ \textbf{Number}}} &
    \multicolumn{3}{c}{\textbf{Train}}&\multicolumn{3}{c}{\textbf{Test}}\\
    \cmidrule(lr{1em}){4-6}\cmidrule(lr{1em}){7-9}
    & & &{\textbf{$1\%$}}&{\textbf{$2\%$}}&{\textbf{$5\%$}}&
    {\textbf{$1\%$}}&{\textbf{$2\%$}}&{\textbf{$5\%$}}\\
\midrule
180 & 5 & 5 & $22.1 \pm 4.3$ & $31.6 \pm 4.1$ & $57.8 \pm 3.4$ & $24.0 \pm 7.0$ & $33.6 \pm 5.8$ & $59.6 \pm 5.2$\\
180 & 5 & 10 & $27.4 \pm 4.1$ & $32.5 \pm 4.1$ & $55.3 \pm 3.6$ & $29.3 \pm 6.6$ & $34.9 \pm 6.0$ & $59.0 \pm 5.9$\\
180 & 10 & 5 & $27.6 \pm 4.3$ & $33.3 \pm 4.2$ & $56.0 \pm 3.5$ & $29.3 \pm 6.6$ & $35.3 \pm 6.0$ & $59.5 \pm 5.9$\\
180 & 10 & 10 & $28.0 \pm 4.3$ & $34.9 \pm 4.8$ & $53.6 \pm 4.4$ & $29.9 \pm 6.4$ & $36.9 \pm 6.6$ & $58.1 \pm 7.5$\\
$\bm{\mathit{200}}$ & $\bm{\mathit{5}}$ & $\bm{\mathit{5}}$ & $40.2 \pm 4.3$ & $55.7 \pm 3.9$ & $\bm{\mathit{69.8}} \pm 3.9$ & $43.5 \pm 5.4$ & $56.9 \pm 5.9$ & $\bm{\mathit{74.2}} \pm 6.3$\\
200 & 5 & 10 & $40.9 \pm 3.8$ & $51.8 \pm 4.1$ & $66.6 \pm 3.4$ & $46.2 \pm 5.5$ & $53.7 \pm 6.5$ & $70.3 \pm 6.3$ \\
200 & 10 & 5 & $41.1 \pm 3.8$ & $50.1 \pm 4.1$ & $64.0 \pm 3.6$ & $45.9 \pm 5.5$ & $52.9 \pm 6.2$ & $68.5 \pm 6.2$\\
200 & 10 & 10 & $37.1 \pm 4.5$ & $46.4 \pm 4.2$ & $60.5 \pm 4.1$ & $41.6 \pm 6.5$ & $50.8 \pm 6.3$ & $65.6 \pm 6.7$\\
$\bm{\mathit{220}}$ & $\bm{\mathit{5}}$ & $\bm{\mathit{5}}$ & $\bm{\mathit{50.9}} \pm 4.8$ & $\bm{\mathit{59.0}} \pm 4.3$ & $66.4 \pm 4.2$ & $\bm{\mathit{51.7}} \pm 7.1$ & $\bm{\mathit{60.7}} \pm 7.0$ & $69.3 \pm 6.8$\\
220 & 5 & 10 & $47.0 \pm 4.8$ & $52.6 \pm 4.6$ & $60.5 \pm 4.2$ & $50.0 \pm 7.5$ & $54.2 \pm 6.9$ & $63.7 \pm 6.9$\\
220 & 10 & 5 & $45.8 \pm 4.8$ & $52.8 \pm 4.6$ & $60.5 \pm 4.2$ & $49.1 \pm 7.3$ & $54.2 \pm 6.9$ & $63.7 \pm 6.9$\\
220 & 10 & 10 &  $41.0 \pm 4.4$ & $43.0 \pm 4.7$ & $47.1 \pm 4.9$ & $45.2 \pm 7.5$ & $47.1 \pm 7.3$ & $51.5 \pm 8.3$\\
240 & 5 & 5 & $47.2 \pm 4.5$ & $50.3 \pm 5.1$ & $54.0 \pm 4.8$ & $49.6 \pm 7.0$ & $52.8 \pm 8.2$ & $57.8 \pm 7.5$\\
240 & 5 & 10 & $42.5 \pm 4.8$ & $43.7 \pm 4.3$ & $50.1 \pm 4.4$ & $46.2 \pm 7.6$ & $47.9 \pm 7.1$ & $53.6 \pm 6.9$\\
240 & 10 & 5 & $42.5 \pm 4.8$ & $43.4 \pm 4.4$ & $47.3 \pm 4.5$ & $46.2 \pm 7.6$ & $47.5 \pm 7.0$ & $51.5 \pm 6.9$\\
240 & 10 & 10 & $40.3 \pm 4.7$ & $41.1 \pm 4.4$ & $41.9 \pm 4.8$ & $44.5 \pm 7.4$ & $45.3 \pm 7.5$ & $48.0 \pm 7.6$  \\
\bottomrule
\end{tabular}
}%
\caption{\textbf{Outlier detection performance by erosion.} We show the percentage of outliers detected when $1\%$, $2\%$ and $5\%$ images are selected by the erosion method for both training and testing set. The true reference outliers for erosion exclude those due to bad positioning, incorrect exposure parameter and heterogeneous pectoral muscle. In total 16 configuration of hyperparameters are tried, and the ones with the optimal performance are in italic and bold. The hyperparameters that have optimal performance in the training set also work the best in the testing set.  \label{tab:performance_erosion}}
\end{table}

\begin{table}
\resizebox{\textwidth}{!}{%
  \begin{tabular}{@{} llll *3c *3c@{}}
    \toprule
    \multirow{2}{*}{\thead{\textbf{Lower}\\ \textbf{Distance}}} & \multirow{2}{*}{\thead{\textbf{Lower}\\ \textbf{Threshold}}}& \multirow{2}{*}{\thead{\textbf{Upper}\\ \textbf{Threshold}}}& \multirow{2}{*}{\thead{\textbf{Hough}\\ \textbf{Threshold}}}&
    \multicolumn{3}{c}{\textbf{Train}}&\multicolumn{3}{c}{\textbf{Test}}\\
    \cmidrule(lr{1em}){5-7}\cmidrule(lr{1em}){8-10}
    & & & &{\textbf{$1\%$}}&{\textbf{$2\%$}}&{\textbf{$5\%$}}&
    {\textbf{$1\%$}}&{\textbf{$2\%$}}&{\textbf{$5\%$}}\\
\midrule
5 & 160 & 180 & 40 & $4.3 \pm 3.1$ & $5.9 \pm 3.1$ & $25.5 \pm 5.2$ & $1.8 \pm 2.9$ & $4.7 \pm 4.9$ & $22.9 \pm 7.2$\\
5 & 160 & 180 & 50 & $10.2 \pm 5.0$ & $15.5 \pm 5.3$ & $22.8 \pm 6.4$ & $8.2 \pm 4.9$ & $12.2 \pm 8.3$ & $20.5 \pm 10.5$\\
5 & 160 & 180 & 60 & $9.6 \pm 4.6$ & $14.7 \pm 6.0$ & $22.5 \pm 6.6$ & $8.6 \pm 6.0$ & $12.6 \pm 8.0$ & $17.9 \pm 9.9$\\
5 & 160 & 220 & 40 & $5.9 \pm 3.7$ & $13.3 \pm 4.9$ & $28.1 \pm 5.5$ & $6.1 \pm 3.8$ & $11.6 \pm 6.6$ & $25.8 \pm 8.3$\\
5 & 160 & 220 & 50 & $13.0 \pm 5.3$ & $14.7 \pm 5.2$ & $22.9 \pm 8.0$ & $9.7 \pm 6.8$ & $11.7 \pm 7.3$ & $22.0 \pm 10.3$\\
5 & 160 & 220 & 60 & $10.9 \pm 4.4$ & $16.2 \pm 5.1$ & $22.7 \pm 6.7$ & $8.8 \pm 5.8$ & $12.8 \pm 8.1$ & $18.7 \pm 9.5$\\
5 & 170 & 180 & 40 & $1.7 \pm 2.6$ & $7.1 \pm 3.8$ & $25.6 \pm 4.9$ & $1.4 \pm 3.0$ & $6.7 \pm 5.4$ & $23.5 \pm 7.1$\\
$\bm{\mathit{5}}$ & $\bm{\mathit{170}}$ & $\bm{\mathit{180}}$ & $\bm{\mathit{50}}$ & $10.0 \pm 3.8$ & $\bm{\mathit{17.3}} \pm 5.3$ & $24.9 \pm 6.6$ & $8.8 \pm 6.5$ & $\bm{\mathit{13.2}} \pm 7.0$ & $22.7 \pm 10.3$\\
5 & 170 & 180 & 60 & $10.9 \pm 4.2$ & $13.8 \pm 5.8$ & $19.8 \pm 7.0$ & $9.1 \pm 6.2$ & $11.5 \pm 7.8$ & $16.4 \pm 8.8$\\
$\bm{\mathit{5}}$ & $\bm{\mathit{170}}$ & $\bm{\mathit{220}}$ & $\bm{\mathit{40}}$ & $7.3 \pm 3.5$ & $13.2 \pm 5.0$ & $\bm{\mathit{32.2}} \pm 5.6$ & $7.1 \pm 5.5$ & $13.6 \pm 8.2$ & $\bm{\mathit{28.8}} \pm 7.7$\\
$\bm{\mathit{5}}$ & $\bm{\mathit{170}}$ & $\bm{\mathit{220}}$ & $\bm{\mathit{50}}$ & $\bm{\mathit{13.3}} \pm 4.6$ & $15.1 \pm 5.4$ & $23.9 \pm 7.4$ & $\bm{\mathit{9.7}} \pm 6.2$ & $12.2 \pm 7.1$ & $19.8 \pm 7.7$\\
5 & 170 & 220 & 60 & $11.5 \pm 4.6$ & $12.6 \pm 5.2$ & $17.0 \pm 6.0$ & $9.1 \pm 6.2$ & $11.8 \pm 7.3$ & $15.6 \pm 8.2$\\
20 & 160 & 180 & 40 & $3.4 \pm 2.1$ &$ 5.8 \pm 3.1$ & $25.0 \pm 5.1$ & $1.5 \pm 2.7$ & $4.4 \pm 6.0$ & $22.9 \pm 7.2$\\
20 & 160 & 180 & 50 & $9.1 \pm 4.3$ & $16.4 \pm 5.4$ & $23.4 \pm 5.9$ & $8.1 \pm 5.1$ & $12.3 \pm 6.9$ & $21.8 \pm 9.5$\\
20 & 160 & 180 & 60 & $9.9 \pm 4.9$ & $15.4 \pm 6.8$ & $22.7 \pm 6.7$ & $7.5 \pm 6.5$ & $12.2 \pm 7.0$ & $17.9 \pm 9.9$\\
20 & 160 & 220 & 40 & $6.2 \pm 3.2$ & $12.3 \pm 4.1$ & $27.7 \pm 5.2$ & $6.6 \pm 5.1$ & $10.7 \pm 5.6$ & $26.2 \pm 7.7$\\
20 & 160 & 220 & 50 & $11.1 \pm 5.0$ & $14.2 \pm 4.6$ & $22.3 \pm 6.0$ & $9.7 \pm 6.1$ & $11.7 \pm 7.3$ & $20.2 \pm 7.7$\\
20 & 160 & 220 & 60 & $11.1 \pm 4.3$ & $16.3 \pm 5.0$ & $22.7 \pm 6.7$ & $9.1 \pm 6.2$ & $12.0 \pm 8.0$ & $17.9 \pm 9.9$\\
20 & 170 & 180 & 40 & $1.7 \pm 2.7$ & $8.3 \pm 4.6$ & $25.2 \pm 4.9$ & $1.7 \pm 3.2$ & $6.6 \pm 5.3$ & $23.9 \pm 7.4$\\
20 & 170 & 180 & 50 & $10.9 \pm 3.7$ & $16.1 \pm 3.8$ & $23.7 \pm 5.5$ & $8.5 \pm 6.4$ & $12.1 \pm 7.3$ & $21.5 \pm 8.4$\\
20 & 170 & 180 & 60 & $10.6 \pm 4.4$ & $13.7 \pm 6.1$ & $19.8 \pm 7.0$ & $9.1 \pm 6.2$ & $11.0 \pm 7.5$ & $16.1 \pm 10.2$\\
20 & 170 & 220 & 40 & $5.8 \pm 3.2$ & $13.9 \pm 4.7$ & $31.1 \pm 5.4$ & $7.3 \pm 5.3$ & $14.5 \pm 7.1$ & $29.2 \pm 7.5$\\
20 & 170 & 220 & 50 & $12.5 \pm 4.6$ & $15.1 \pm 5.2$ & $23.5 \pm 8.1$ & $10.6 \pm 6.6$ & $12.4 \pm 7.3$ & $20.1 \pm 8.4$\\
20 & 170 & 220 & 60 & $11.1 \pm 4.2$ & $12.5 \pm 5.4$ & $17.6 \pm 6.0$ & $9.7 \pm 6.1$ & $10.8 \pm 7.2$ & $14.9 \pm 8.9$\\
\bottomrule
\end{tabular}
}%
\caption{\textbf{Outlier detection performance by pectoral muscle analysis.} We show the percentage of outliers detected when $1\%$, $2\%$ and $5\%$ images are selected by the pectoral muscle analysis for both training and testing set. The true reference outliers for pectoral muscle analysis only include those with heterogeneous pectoral muscle regions. In total 24 configuration of hyperparameters are tried, and the ones with the highest performance in the training set are in italic and bold. The hyperparameters with optimal performance in the training set also have satisfactory performance in the testing set. \label{tab:performance_muscle}}
\end{table}